\documentclass[11pt,a4paper,fleqn]{article}

\usepackage{amsmath,amssymb,amsthm,enumerate,cite}%,backref

\newcommand{\vspacebefore}{\raisebox{0ex}[2.5ex][0ex]{\null}}
\newcommand{\vspacebeforemore}{\raisebox{0ex}[3.3ex][0ex]{\null}}
\newcommand{\p}{\partial}
\newcommand{\const}{\mathop{\rm const}\nolimits}
\newcommand{\arccosh}{\mathop{\rm arccosh}\nolimits}
\newcommand{\arcsinh}{\mathop{\rm arcsinh}\nolimits}
\newcommand{\Equiv}{\mathop{ \sim}}

\newcommand{\sign}{\mathop{\rm sign}\nolimits}

\newcounter{tbn}
\newcounter{tabul}
\newcounter{casetran}
\newcounter{mcasenum}

{\theoremstyle{definition}

\newtheorem{note}{Note}
\begin{document}

\par\noindent {\LARGE\bf
Exact Solutions of Diffusion--Convection Equations%\\
%Brief History Review
\par}
{\vspace{6mm}\par\noindent {\bf\large Nataliya M. Ivanova} \par\vspace{3mm}\par}
{\vspace{3mm}\par\noindent {\it
Institute of Mathematics of NAS of Ukraine, 3 Tereshchenkivska Str.,
01601 Kyiv-4, Ukraine\\
% Department of Mathematics, University of British Columbia, Vancouver, BC, V6T 1Z2, Canada\\
}}
{\noindent {\it
e-mail: ivanova@imath.kiev.ua
} \par}

{\vspace{5mm}\par\noindent\hspace*{8mm}\parbox{140mm}{\small
In the presented paper known (up to the beginning of 2008) Lie- and non-Lie exact solutions
of different $(1+1)$-dimensional diffusion--convection equations of form
$f(x)u_t=(g(x)A(u)u_x)_x+h(x)B(u)u_x$ are collected.
%All possible Lie reductions of equations under considerations are performed and new exact solutions are constructed.
}\par\vspace{5mm}}

%\tableofcontents

%\newpage

\section{Introduction}\label{Intro}

This is a review paper, where we present a brief summary of known exact solutions
of variable coefficient $(1+1)$-dimensional diffusion--convection equations of form
\begin{equation} \label{eqDKfgh}
f(x)u_t=(g(x)A(u)u_x)_x+h(x)B(u)u_x,
\end{equation}
where $f=f(x),$ $g=g(x),$ $h=h(x),$ $A=A(u)$ and $B=B(u)$ are arbitrary smooth functions of their variables,
$f(x)g(x)A(u)\!\neq\! 0.$

Our aim is not to give a physical interpretation of the solution of diffusion equations
(that is too huge and cannot be reached in the scope of a short paper),
but to list the already known exact solutions of equations from the class under consideration.
However, in some cases we give a short discussion of the nature of the listed solutions.

The majority of the listed solutions have been obtained by means of different symmetry methods, such as reduction
with respect to Lie and non-Lie symmetries, separation of variables, equivalence transformations, etc.

Let us note that the constant coefficient diffusion equations ($f=g=1$, $B=0$) are well investigated
and some of exact solutions
given below were summarized before in~\cite{Polyanin&Zaitsev2004,Ibragimov1994V1}.

Our paper is organized as follows. First of all we adduce solutions of the linear heat equation obtained
by means of various symmetry methods.
In Section~\ref{SectionOnSolutionOfLinearizEq} the linearizable Burgers, Fujita--Storm and Fokas--Yortsos equations
are considered.
Lie reduction of constant coefficient nonlinear diffusion equation ($hB=0$, $f=g=1$)
is performed in Section~\ref{SectionOnSolutionsOfNDEGenCase}.
Solutions of constant coefficient diffusion equations with exponential nonlinearity
are adduced in Section~\ref{SectionOnSolutionsOfNDEexpu}.
Solutions of constant coefficient diffusion equations with power nonlinearity are presented
in Section~\ref{SectionOnSolutionsOfNDEumu}.
The important particular case of such equations, namely, the fast diffusion equation, is studied in more detail
in Section~\ref{SectionOnSolutionOfFastDifEq}.
Diffusion equations with other nonlinearities are briefly discussed in Section~\ref{SectionOnSolutionsOfNDEMisc}.
The next considered case (Section~\ref{SectionOnSolutionsOfNCCDCE}) covers the nonlinear constant coefficient
diffusion--convection equations ($f=g=h=1$).
In Section~\ref{SectionRadSym} we adduce a brief analysis of known solutions of $n$-dimensional radially symmetric nonlinear diffusion equations.
In Section~\ref{SectionOnSolutionsOfNVCDCE} exact solutions of some variable coefficient
diffusion--convection equations are collected.
At last, in Sections~\ref{SectionOnExactSolOfGenBurEq} and~\ref{SectionOnSolutionOfSL2REq}
we present a detailed analysis of interesting
variable coefficient equations having distinguished invariance properties.

In the Appendix~\ref{AppendixGroupClassif} we adduce the complete results of group classification
of equations~\eqref{eqDKfgh}
with respect to the extended group~$\hat G^{\sim}$ of equivalence transformations~\eqref{EquivGroup}.
%Appendix~\ref{AppendixLieReductionsOfVarCoefEqs} contains reductions of variable coefficient equations~\eqref{eqDKfgh}
%with respect to the inequivalent subalgebras of the corresponding Lie symmetry algebras.

Below, if it is not indicated separately, $\alpha$, $\varepsilon_i$, $\lambda$, $a$, $b$, $c$, $c_i$
are arbitrary constants, $\varepsilon=\pm1$.
For convenience we use double numeration T.N of classification cases and local equivalence
transformations, where T denotes the number of table and N does the number of case (or transformation,
or solution) in table~T.
The notion ``equation~T.N'' is used for the equation of form~\eqref{eqDKfgh} where
the parameter-functions $f$, $g$, $h$, $A$, $B$ take values from the corresponding case.

\section{Linear heat equation}
Systematical investigation of invariant solutions of different diffusion equations was started
by the case of linear heat equation~\cite{Katkov1965,Katkov1968,Olver1986,Ovsiannikov1959,Ovsiannikov1966,Miller1977}
\mathversion{bold}\begin{equation}\label{eqLHE}
u_t=u_{xx}
\end{equation}\mathversion{normal}%
which is invariant with respect to the six dimensional symmetry algebra generated by the vector fields
\begin{gather*}
Q_1=\p_x, \qquad Q_2=\p_t, \qquad Q_3=u\p_u,\qquad Q_4=2t\p_t+x\p_x,\\
Q_5=2t\p_x-xu\p_u, \qquad Q_6=4t^2\p_t+4tx\p_x-(x^2+2t)u\p_u
\end{gather*}
(For the moment we are ignoring the trivial infinite-dimensional subalgebras coming from the linearity
of the heat equation
and corresponding to the linear superposition principle).

The most general solution obtainable from a given solution $u=f(t,x)$ by group transformations is of the form
\begin{equation}\label{GroupActionLHE}
\tilde u=\frac{\varepsilon_3}{\sqrt{1+4\varepsilon_6t}}
e^{-\frac{\varepsilon_5x+\varepsilon_6x^2-\varepsilon_5^2t}{1+4\varepsilon_6t}}
f\left(\frac{\varepsilon_4^2t}{1+4\varepsilon_6t}-\varepsilon_2,
\frac{\varepsilon_4(x-2\varepsilon_5t)}{1+4\varepsilon_6t}-\varepsilon_1\right)+v(t,x),
\end{equation}
where
%$\varepsilon_1,\ldots,\varepsilon_6$ are arbitrary constants and
$v(t,x)$ is an arbitrary solution to the linear heat equation~\cite{Olver1986}.
Considering the higher-order symmetry generators, one can prove that if $u=f(t,x)$
is a solution of the linear heat equation than
\begin{gather*}
\tilde u=2tf_x(t,x)+xf(t,x)\quad\mbox{and}\\
\tilde u=t^2f_t(t,x)+txf_x(t,x)+\frac14(2t+x^2)f(t,x)
\end{gather*}
are also solutions of the same equation~\cite{Anderson&Ibragimov1979}.

All possible inequivalent (with respect to inner automorphisms)
one-dimensional subalgebras of the given algebra are exhausted by the ones
listed in Table~\ref{TableReductionForLHE}~\cite{Weisner1959} together with
the corresponding ansatzes and the reduced ODEs.

\setcounter{tbn}{0}
{\begin{center}\refstepcounter{tabul}\label{TableReductionForLHE}
Table~\thetabul. Reduced ODEs for linear heat equation~\eqref{eqLHE}
\\[1.5ex] \footnotesize
\setcounter{tbn}{0}
\renewcommand{\arraystretch}{1.2}
\begin{tabular}{|l|l|c|c|l|}
\hline \vspacebefore
N&Subalgebra& Ansatz $u=$& $\omega$ &\hfill {Reduced ODE\hfill} \\
\hline
\refstepcounter{tbn}\thetbn&$\langle Q_4+aQ_3\rangle$ & $t^a\varphi(\omega)$ & $x/\sqrt{t}$
   & $\varphi''+\omega\varphi'/2-a\varphi=0$\\
\refstepcounter{tbn}\thetbn&$\langle Q_2+Q_6+aQ_3\rangle$
    & $(4t^2+1)^{1/4}\varphi(\omega)e^{-(4t^2+1)^{-1}tx^2-a\arctan(2t)/2}$
      & $(4t^2+1)^{-1/2}x$ & $\varphi''+(a+\omega^2)\varphi=0$\\
\refstepcounter{tbn}\thetbn&$\langle Q_2-Q_5\rangle$ & $\varphi(\omega)e^{tx+2t^3/3}$ & $x+t^2$
    & $\varphi''=\omega\varphi$\\
\refstepcounter{tbn}\thetbn&$\langle Q_2+aQ_3\rangle$ & $\varphi(\omega)e^{at}$ & $x$ & $\varphi''=a\varphi$\\
\refstepcounter{tbn}\thetbn&$\langle Q_1\rangle$ & $\varphi(\omega)$ & $t$ &
$\varphi'=0$\\
\refstepcounter{tbn}\thetbn&$\langle Q_3\rangle$ & --- & --- & \hfill {---\hfill} \\
\hline
\end{tabular}
\end{center}}

Thus we have the following solutions of~(\ref{eqLHE}):
\begin{gather*}
u=t^ae^{-x^2/(8t)}\left(c_1U\Big(2a+\frac12,\frac x{\sqrt{2t}}\Big)+c_2V\Big(2a+\frac12,\frac x{\sqrt{2t}}\Big)\right),\\
u=(4t^2+1)^{1/4}\left(c_1W\Big(-\frac a2,\frac x{\sqrt{8t^2+2}}\Big)
   +c_2W\Big(-\frac a2,-\frac x{\sqrt{8t^2+2}}\Big)\right)e^{-\frac{tx^2}{(4t^2+1)}-\frac a2\arctan(2t)},\\
u=c_1e^{\alpha^2 t}\cosh(\alpha x+c_2),\quad u=c_1x+c_2,\quad u=c_1e^{-\alpha^2 t}\cos(\alpha x+c_2),
\end{gather*}
where $U(b,z)$, $V(b,z)$, $W(c,z)$ are parabolic cylinder functions~\cite{Abramowitz&Stegun1970}.

Multiplicative separation of variables leads to the solutions invariant with respect to $\langle Q_2+aQ_3\rangle$.
The additive separation of variables yields an exact solution of form
\[
u=c_1x^2+c_2x+2c_1t.
\]

The known $Q$-conditional symmetry operators and corresponding reductions
are adduced in Table~\ref{TableNonclasReductionForLHE}.~\cite{Fushchych&Serov&Popovych1992}

\begin{center}\refstepcounter{tabul}\label{TableNonclasReductionForLHE}
Table~\thetabul. Nonclassical reductions for linear heat equation~\eqref{eqLHE}
\\[1.5ex] \footnotesize
\begin{tabular}{|c|c|c|c|}
\hline
&&&\\[-3mm]
N & Operator $Q$  & Ansatz & Reduced equation \\[1mm]
\hline
&&&\\[-3mm]
1  &  $-x\p_t+\p_x$  &  $u=\varphi\left(t+\frac{x^2}{2}\right)$ &
$\varphi''=0$ \\[1mm]
2  &  $-x\p_t+\p_x+x^3\p_u$  &
$u=\varphi\left(t+\frac{x^2}{2}\right)+\frac{x^4}{4}$ &
$\varphi''= -3$ \\[1mm]
3  &  $x^2\p_t-3x\p_x-3u\p_u$  &  $u=x\varphi\left(t+
\frac{x^2}{6}\right)$ & $ \varphi''=0 $ \\[1mm]
4  &  $x^2\p_t-3x\p_x-(3u+x^5)\p_u$  &  $u=x\varphi\left(t+
\frac{x^2}{6}\right)+\frac{x^5}{12}$ & $ \varphi''=-15 $ \\[1mm]
5  &  $x\p_x+u\p_u$  &  $u=x\varphi(t)$  &  $\varphi'=0$ \\[1mm]
6 & $\coth x\p_x+u\p_u$ & $u=\varphi(t)\cosh x$ & $\varphi'-\varphi=0$ \\[1mm]
7  &  $-\cot x\p_x+u\p_u$ & $u=\varphi(t)\cos x$ & $\varphi'+ \varphi=0$ \\[1mm]
8  &  $\p_x-u\p_u-\frac{u}{2t-x}\p_u$  & $u=(2t-x)e^{-x}
\varphi(t)$  &  $\varphi'-\varphi=0$ \\[1mm]
9 & $\p_x-\sqrt{-2(t+u)}\p_u$ &
$u=-t-\frac{1}{2}[x+\varphi(t)]^2$  &  $\varphi'=0$ \\[1mm]
10  &  $\left(t+\frac{x^2}{2}\right)\p_t-x\p_x$  &
$u=\varphi\left(tx+\frac{x^2}{3!}\right)$  &  $\varphi''=0$  \\[1mm]
\hline
\end{tabular}
\end{center}

If function $f(t,x)$ is an arbitrary solution of the linear heat
equation and $u$ is the general integral of the ODE
$f_x dt+fdx=0$,
then $u$ satisfies the linear heat equation. This statement
can be considered as another algorithm of generating solutions of
the linear heat equation~\cite{Fushchych&Serov&Popovych1992}. Indeed, even starting from a rather trivial solution of
the heat equation $u=1$ one gets the chain of quite interesting solutions
\[
1 \to x \ \to \ t + \frac{x^2}{2!} \ \to \ t x+\frac{x^3}{3!} \ \to
\ \cdots ,
\]
and among them the solutions
\begin{gather*}
\frac{x^{2m}}{(2m)!}+\frac{t}{1!} \frac{x^{2m-2}}{(2m-2)!}+
\frac{t^2}{2!} \frac{x^{2m-4}}{(2m-4)!}+\cdots +
\frac{t^{m-1}}{(m-1)!}\frac{x^2}{2!}+\frac{t^m}{m!},
\\[1ex]
 \frac{x^{2m+1}}{(2m+1)!}+\frac{t}{1!}
\frac{x^{2m-1}}{(2m-1)!}+\frac{t^2}{2!}
\frac{x^{2m-3}}{(2m-3)!}+ \cdots +
\frac{t^{m-1}}{(m-1)!}\frac{x^3}{3!}+\frac{t^m}{m!} \frac{x}{1!},
\end{gather*}
called often the {\em heat polynomials}~\cite{Widder1975}.

\section{Linearizable equations}\label{SectionOnSolutionOfLinearizEq}

Class~\eqref{eqDKfgh} contains three equations, namely
Burgers equation
\mathversion{bold}\begin{equation}\label{eqBurgE}
u_t=u_{xx}+2uu_x,
\end{equation}
Fujita--Storm equation
\begin{equation} \label{eqFuSE}
u_t=\left(u^{-2}{u_x}\right)_x
\end{equation}
and Fokas--Yortsos equation
\begin{equation}\label{eqFYE}
u_t=(u^{-2}u_x)_x+u^{-2}u_x,
\end{equation}\mathversion{normal}
that are linearizable by
the potential equivalence hodograph transformation and additional local equivalence
transformations~\cite{Bluman&Kumei1980,Storm1951,Forsyth1906,Hopf1950,Cole1951,
Fokas&Yortsos1982,Strampp1982b,Popovych&Ivanova2005PETs,LisleDissertation}:

{\noindent\unitlength=1mm\fboxsep=2mm
\begin{picture}(160,58)
\put(0,2){\line(0,1){14}}\put(0,2){\line(1,0){54}}\put(54,2){\line(0,1){14}}\put(0,16){\line(1,0){54}}
\put(0,8){{
\parbox{52mm}{\hfil$\check u_{\check t}=(\check u^{-2}\check u_{\check x})_{\check x}$\hfil \\[1ex]
\mbox{}\hfil$\check v_{\check x}=\check u,$
$\check v_{\check t}=\check u^{-2}\check u_{\check x}$\hfil } }}
\put(0,39){\line(0,1){14}}\put(0,39){\line(1,0){54}}
\put(54,39){\line(0,1){14}}\put(0,53){\line(1,0){54}}
\put(0,45){{
\parbox{52mm}{\hfil$\hat u_{\hat t}=
(\hat u^{-2}\hat u_{\hat x})_{\hat x}+\hat u^{-2}\hat u_{\hat x}$\hfil \\[1ex]
\mbox{}\hfil$\hat v_{\hat x}=\hat u,$ $\hat v_{\hat t}=\hat u^{-2}\hat u_{\hat x}-\hat u^{-1}$\hfil } }}
\put(106,39){\line(0,1){14}}\put(106,39){\line(1,0){54}}
\put(160,39){\line(0,1){14}}\put(106,53){\line(1,0){54}}
\put(106,45){ {
\parbox{50mm}{ \hfil$\tilde u_{\tilde t}=
\tilde u_{\tilde x\tilde x}+2\tilde u\tilde u_{\tilde x}$\hfil \\[1ex]
\mbox{}\hfil$\tilde v_{\tilde x}=\tilde u,$
$\tilde v_{\tilde t}=\tilde u_{\tilde x}+\tilde u^2$\hfil } }}
\put(106,2){\line(0,1){14}}\put(106,2){\line(1,0){54}}
\put(160,2){\line(0,1){14}}\put(106,16){\line(1,0){54}}
\put(106,8){ {
\parbox{50mm}{ \hfil$u_t=u_{xx}$\hfil \\[1ex]
\mbox{}\hfil$v_x=u,$ $v_t= u_x$\hfil } }}
\put(54,9){\vector(1,0){52}}\put(106,9){\vector(-1,0){52}}
\put(54,11){\parbox{52mm}{\hfil $\check t=t,$ $\check x=v,$ $\check u=u^{-1},$ $\check v=x$\hfil }}
\put(54,46){\vector(1,0){52}}\put(106,46){\vector(-1,0){52}}
\put(54,48){\parbox{52mm}{\hfil $\hat t=\tilde t,$ $\hat x=\tilde v,$ $\hat u=\tilde u^{-1},$
$\hat v=\tilde x$\hfil }}
\put(27,24){\vector(0,-1){8}}\put(27,31){\vector(0,1){8}}
\put(0,27){\parbox{54mm}{\hfil $\check t=\hat t,$ $\check x=e^{\hat x},$
$\check u=e^{-\hat x}\hat u,$ $\check v=\hat v$\hfil }}
\put(133,24){\vector(0,-1){8}}\put(133,31){\vector(0,1){8}}
\put(106,27){\parbox{54mm}{\hfil $t=\tilde t,$ $x=\tilde x,$ $u=\tilde ue^{\tilde v},$
$v=e^{\tilde v}$\hfil }}
\put(54,26.55){\parbox{52mm}{\hfill $\displaystyle\tilde u=\frac{v_x}v,
\ v_t=v_{xx}\ \Longleftarrow$\hspace*{2mm} }}
\end{picture}}
\smallskip

Therefore, applying the above transformations to the well-known solutions of the linear heat equation
one can easily construct solutions
of the linearizable equations.

Thus, e.g., the fundamental (source) solution $u=(4\pi t)^{-1/2}e^{-x^2/(4t)}$ and
dipole solution $u=-((4\pi t)^{-1/2}e^{-x^2/(4t)})_x$ are mapped into the separable and self-similar solutions
of the Fujita--Storm equation~\cite{Polyanin&Zaitsev2004}
\begin{gather*}
u=(4\pi t)^{1/2}e^{v^2}\quad\mbox{where}\quad x=\pi^{-1/2}\int_0^ve^{-y^2}dy,\\
\mbox{and}\quad
u=x^{-1}(2t)^{1/2}\left(\ln\dfrac{1}{4\pi tx^2}\right)^{-1/2}.
\end{gather*}
correspondingly.

Other solutions of the linear heat equation presented in the previous section
yield the following explicite exact solutions of the Fujita--Storm equation~\cite{Ivanova&Popovych&Sophocleous2007Part4}:
\begin{gather*}
%u=c,\quad u=\pm\frac1{\sqrt{c_1(x-2t)+c_0}},\quad u=\pm\frac{1}{\sqrt{x^2\pm e^{2t}}},
%\quad u=\pm\frac{1}{\sqrt{e^{-2t}\pm x^2}},\\
u=c,\quad u=x^{-1},\quad u=(x-2t)^{-1/2},\quad u=(x^2\pm e^{2t})^{-1/2},
\quad u=\pm(e^{-2t}- x^2)^{-1/2},\\
u=\frac1{4\sqrt{24t^2+x}\sqrt{-6t\pm\sqrt{24t^2+x}}},\quad u=\frac t{x\sqrt{-t\ln(x\sqrt t)}},\\
u=\frac1{\sqrt{c_1^2e^{-2t}+2e^{-8t}+2e^{-4t}x}\sqrt{4-e^{8t}(-c_1e^{-t}\pm\sqrt{c_1^2e^{-2t}+2e^{-8t}+2e^{-4t}x})^2}},\\
u=\frac1{\sqrt{c_1^2e^{2t}+2e^{8t}+2e^{4t}x}\sqrt{e^{-8t}(-c_1e^{-t}\pm\sqrt{c_1^2e^{-2t}+2e^{-8t}+2e^{-4t}x})^2-4}}.
\end{gather*}

In~\cite{Tychynin&Petrova&Tertyshnyk2007} the following formula for deriving exact solutions
of the Fujita--Storm equation is derived: if $u(t,x)$ is a solution of the Fujita--Storm equation~\eqref{eqFuSE}
then
\[
v=u+\frac{2tu_t+u+xu_x}{-t(u_x-\frac12u^2)-\frac x2u}
\]
is also solution of the same equation.

Similarly~\cite{Tychynin&Rasin2004}, if $u(t,x)$ is a solution of the and the Fokas--Yortsos equation~\eqref{eqFYE}
then
\[
v=\frac{u^2}{u-u_x}
\]
is also solution of the same equation.
In~\cite{Tychynin&Rasin2004} a new exact solution of the Fokas--Yortsos equation is presented:
\[
u=\frac12\frac{W(e^{4t+x})}{1+W(e^{4t+x})},
\]
where $W(x)$ is the Lambert $W$ function determined as $W(x)e^{W(x)}=x$.
Let us note that other exact solutions of the Fokas--Yortsos equation~\eqref{eqFYE} can be easily
recovered from the solution set of the Fujita--Storm equation~\eqref{eqFuSE} by means of
application of the local transformation of variables shown in the above scheme.

The same tricks can be used for obtaining exact solutions of the remaining linearizable equations.
However, since the adduced transformations are nonlocal, sometimes it could be easier to search directly for solutions
of the nonlinear equations.
Thus, e.g., one can easily find Lie solutions of the Burgers equation:
\begin{gather*}
u=\frac{c_1-x}{2t+c_2},\quad u=\varepsilon+\frac1{x+2\varepsilon t+c},\quad u=\frac{2x+c_1}{x^2+c_1x+2t+c_2},\\
u=\frac{6(x^2+2t+c_1)}{2x^3+12tx+6c_1t+c_2},\quad u=\dfrac{c_1}{1+c_2e^{-c_1^2t-c_1x}},\quad
u=-\varepsilon+\frac a2\frac{e^{c_1(x-2\varepsilon t)}-c_2}{e^{c_1(x-2\varepsilon t)}+c_2},\\
u=-\varepsilon+c_1\tanh(c_1(x-2\varepsilon t)+c_2),\quad
u=\frac\lambda{2(\lambda^2t+c_1)}\left(2\tanh\frac{\lambda x+c_2}{\lambda^2t+c_1}-\lambda x-c_2\right),\\
u=-\varepsilon-c_1\tan(c_1(x-2\varepsilon t)+c_2),
\quad u=\frac{\lambda\cos(\lambda x+c_1)}{c_2e^{\lambda^2t}+\sin(\lambda x+c_1)},\\
u=\frac{c_1}{\sqrt{\pi(t+c_2)}}\exp\left(-\frac{x+c_3}{4(t+c_2)}\right)
\left(c_1\mathrm{erf}\frac{x+c_3}{2\sqrt{t+c_2}}+c_4\right)^{-1},\\
u=-\frac{\cos 2x e^{-3t}+c_1\sin x}{-\cos x\sin x e^{-3t}+c_1\cos x+c_2e^t},\quad
u=-\frac{c_1e^{-t}(\cos x+\sin x)+c_2e^{t+x}}{c_1e^{-t}(\cos x-\sin x)-c_2e^{t+x}+b}.
\end{gather*}
Here $\mathrm{erf}z=\frac2{\sqrt\pi}\int_0^ze^{-\xi^2}d\xi$ is the error function
also called the probability integral. The last two solutions were found in~\cite{QinMeiFan2007}.

Solutions of the Fujita--Storm and Fokas--Yortsos equations can be singled out form the solutions adduced
in Sections~\ref{SectionOnSolutionsOfNDEumu} and~\ref{SectionOnSolutionsOfNCCDCE} taking $\mu=-2$, $\nu=-2$.
(Note that all these solutions can be also reconstructed from ones of the linear heat equation by means of
potential equivalence transformations.)

If $u(t,x)$ is a solution of the Burgers equation~\eqref{eqBurgE}
then
\[
v=u+\frac{u_t}{u_x+u^2}, \quad u=u+\frac{u_t+u_x}{u_x+u^2+u}
\]
are also solutions of the same equation~\cite{Tychynin&Petrova&Tertyshnyk2007}.

Potential equivalence transformations were used to obtain solutions of some boundary-value problems adduced
in~\cite{Bluman&Kumei1980,Strampp1982b,Storm1951,Munier&Burgan&Gutierres&Fijalkow&Feix1981}

\section{Nonlinear diffusion equations. General case}\label{SectionOnSolutionsOfNDEGenCase}

Consider now the class of nonlinear diffusion equations
\mathversion{bold}\begin{equation}\label{eqNDE}
u_t=(A(u)u_x)_x,
\end{equation}\mathversion{normal}%
where $A_u\ne0$. Lie symmetries of this class have been studied in~\cite{Ovsiannikov1959}.
The Lie symmetry algebra of equation from class~\eqref{eqNDE} with arbitrary value
of parameter-function~$A(u)$ is three-dimensional and spanned by
\[
Q_1=\p_t,\quad Q_2=\p_x,\quad Q_3=2t\p_t+x\p_x.
\]

Taking into account discrete symmetry transformations of changing sings of independent variables
one can formulate the following statement.
If $u=f(t,x)$ is a solution of equation~\eqref{eqNDE},
then $\tilde u=f(\varepsilon_1^2t+\varepsilon_2,\varepsilon_1x+\varepsilon_3)$
is also solution of the same equation.

All possible inequivalent (with respect to inner automorphisms)
one-dimensional subalgebras of the given symmetry algebra,the corresponding ansatzes and the reduced ODEs
are exhausted by the ones listed in Table~\ref{TableReductionForNDE}.

{\begin{center}\refstepcounter{tabul}\label{TableReductionForNDE}
Table~\thetabul. Reduced ODEs for nonlinear diffusion equations~\eqref{eqNDE}, $A_u\ne0$.
\\[1.5ex] \footnotesize
\setcounter{tbn}{0}
\renewcommand{\arraystretch}{1.2}
\begin{tabular}{|l|l|c|c|l|}
\hline \vspacebefore
N&Subalgebra& Ansatz $u=$& $\omega$ &\hfill {Reduced ODE\hfill} \\
\hline
\refstepcounter{tbn}\thetbn&$\langle Q_1\rangle$ & $\varphi(\omega)$ & $x$ & $(A(\varphi)\varphi')'=0$\\
\refstepcounter{tbn}\thetbn&$\langle Q_2\rangle$ & $\varphi(\omega)$ & $t$ & $\varphi'=0$\\
\refstepcounter{tbn}\thetbn&$\langle Q_1+\varepsilon Q_2\rangle$ & $\varphi(\omega)$
    &$x-\varepsilon t$ &$\varepsilon\varphi'=-(A(\varphi)\varphi')'$\\
\refstepcounter{tbn}\thetbn&$\langle Q_3\rangle$ & $\varphi(\omega)$ & $x/\sqrt{t}$
    & $\omega\varphi'=-2(A(\varphi)\varphi')'$\\
\hline
\end{tabular}
\end{center}}
The first three equations can be easily integrated for all values of $A(u)$.
Solutions of the last equation are known for many functions $A(u)$ (see Section~\ref{SectionOnSolutionsOfNDEMisc}).

Consider now in more detail equations with wider symmetry algebras.
Up to the group of equivalence transformations
\[
\tilde t=\varepsilon_1t+\varepsilon_4,\quad \tilde x=\varepsilon_2x+\varepsilon_5,\quad
\tilde u=\varepsilon_3u+\varepsilon_6,\quad \tilde A=\varepsilon_1^{-1}\varepsilon_2^2A
\]
there exist three inequivalent cases of extensions of Lie symmetry algebra~\cite{Ovsiannikov1959}:
 $a=e^u$, $a=u^\mu$, $\mu\ne-4/3$ and $a=u^{-4/3}$.

\section{Nonlinear diffusion equations. Exponential nonlinearity}\label{SectionOnSolutionsOfNDEexpu}

We start from the equation with exponential nonlinearity
\mathversion{bold}\begin{equation}\label{solv1}
u_t=(e^uu_x)_x,
\end{equation}\mathversion{normal}
having the four-dimensional Lie algebra spanned by the operators
\[
Q_1=\p_t,\ Q_2=t\p_t-\p_u,\ Q_3=\p_x,\ Q_4=x\p_x+2\p_u.
\]
The only non-zero commutators of these operators are
$[Q_1,Q_2]=Q_1$ and $[Q_3,Q_4]=Q_3.$ Therefore $A^{\max}$ is a realization
of the algebra~$2A_{2.1}$~\cite{Mubarakzyanov1963}.
All the possible inequivalent (with respect to inner automorphisms)
one-dimensional subalgebras of~$2A_{2.1}$~\cite{Patera&Winternitz1977} are exhausted by the ones
listed in Table~\ref{TableReductionForEu}.

{\begin{center}\refstepcounter{tabul}\label{TableReductionForEu}
Table~\thetabul. Reduced ODEs for (\ref{solv1}). $\alpha\ne0,\ \varepsilon=\pm1,\ \delta=\sign t$.
\\[1.5ex] \footnotesize
\begin{tabular}{|l|l|c|c|l|}
\hline \vspacebefore
N&Subalgebra& Ansatz $u=$& $\omega$ &\hfill {Reduced ODE\hfill} \\
\hline
1&$\langle Q_3\rangle$ & $\varphi(\omega)$ & $t$ & $\varphi'=0$\\
2&$\langle Q_4\rangle$ & $\varphi(\omega)+2\ln|x|$ & $t$ & $\varphi'=2e^{\varphi}$\\
3&$\langle Q_1\rangle$ & $\varphi(\omega)$ & $x$ & $(e^{\varphi})''=0$\\
4&$\langle Q_2\rangle$ & $\varphi(\omega)-\ln|t|$ & $x$ & $(e^{\varphi})''=-\delta$\\
5&$\langle Q_1+\varepsilon Q_3\rangle$ & $\varphi(\omega)$ & $x-\varepsilon t$ &
$(e^{\varphi})''=-\varepsilon\varphi'$\\
6&$\langle Q_2+\varepsilon Q_3\rangle$ & $\varphi(\omega)-\ln|t|$ & $x-\varepsilon\ln|t|$
& $(e^{\varphi})''=-\delta(\varepsilon\varphi'+1)$\\
7&$\langle Q_1+\varepsilon Q_4\rangle$ & $\varphi(\omega)+2\varepsilon t$ & $xe^{-\varepsilon t}$ &
$(e^{\varphi})''=-\varepsilon\omega\varphi'+2\varepsilon$\\
8&$\langle Q_2+\alpha Q_4\rangle$ & $\varphi(\omega)+(2\alpha-1)\ln|t|$ & $x|t|^{-\alpha}$ &
     $(e^{\varphi})''=\delta(-\alpha\omega\varphi'+2\alpha-1)$\\
\hline
\end{tabular}
\end{center}

Solving the equations \ref{TableReductionForEu}.1--\ref{TableReductionForEu}.5
we have the following solutions of~(\ref{solv1}):
\[
u=\ln|c_1x+c_0|, \qquad
u=\ln\left(\dfrac{-x^2}{2t}+\dfrac{c_1x+c_0}t\right),
\]
\[
u=\varphi(x-\varepsilon t)\quad \mbox{where}\
\int{\frac{e^{\varphi}}{c_1-\varepsilon\varphi}}d\varphi=x-\varepsilon t+c_0.
\]

Equation~(\ref{solv1}) admits an additive separation of variable that leads to the solution invariant
with respect to scale transformation.

If $u=f(t,x)$ is a solution of equation~\eqref{solv1}, then
\[
\tilde u=f(\varepsilon_1t+\varepsilon_3,\varepsilon_2x+\varepsilon_4)-\varepsilon_1+2\varepsilon_2
\]
is also solution of the same equation.

\section{Nonlinear diffusion equations. Power nonlinearities}\label{SectionOnSolutionsOfNDEumu}

Another case of equations admitting extension of the Lie symmetry algebra is the one having power nonlinearity
\mathversion{bold}\begin{equation} \label{solv5}
u_t=\left(|u|^{\mu}u_x\right)_x.
\end{equation}\mathversion{normal}
As in the previous cases the invariance algebra of~(\ref{solv5})
\[
A^{\max}=\langle Q_1=\p_t,\ Q_2=t\p_t-\mu^{-1}u\p_u,\ Q_3=\p_x,\ Q_4=x\p_x+2\mu^{-1}u\p_u\rangle
\]
is a realization of the algebra~$2A_{2.1}$.

If $u=f(t,x)$ is a solution of equation~\eqref{solv5}, then
\[
\tilde u=\varepsilon_1^{-1}\varepsilon_2^2f(\varepsilon_1^\mu t+\varepsilon_3,\varepsilon_2^\mu x+\varepsilon_4)
\]
is also solution of the same equation.

The result of reduction~(\ref{solv5})
under inequivalent subalgebras of~$A^{\max}$ is written down in Table~\ref{TableReductionFormuforall}.

\begin{center}\refstepcounter{tabul}\label{TableReductionFormuforall}
Table~\thetabul. Reduced ODEs for (\ref{solv5}). $\mu\ne0$ $\alpha\ne0,$ $\varepsilon=\pm1,$ $\delta=\sign t.$
\\[1.5ex] \footnotesize
\renewcommand{\arraystretch}{1.2}
\begin{tabular}{|l|l|c|c|l|}
\hline \vspacebefore
N&Subalgebra& Ansatz $u=$& $\omega$ &\hfill {Reduced ODE\hfill} \\
\hline
1&$\langle Q_3\rangle$ & $\varphi(\omega)$ & $t$ & $\varphi'=0$\\
2&$\langle Q_4\rangle$ & $\varphi(\omega)|x|^{2/\mu}$ & $t$ &
   $\varphi'=2\mu^{-2}(2+\mu)\varphi^{\mu+1}$\\
3&$\langle Q_1\rangle$ & $\varphi(\omega)$ & $x$ & $(\varphi^{\mu}\varphi')'=0$\\
4&$\langle Q_2\rangle$ & $\varphi(\omega)|t|^{-1/\mu}$ & $x$
  & $(\varphi^{\mu}\varphi')'=-\delta\mu^{-1}\varphi$\\
5&$\langle Q_1+\varepsilon Q_3\rangle$ & $\varphi(\omega)$ & $x-\varepsilon t$ &
     $(\varphi^{\mu}\varphi')'=-\varepsilon\varphi'$\\
6&$\langle Q_2+\varepsilon Q_3\rangle$ & $\varphi(\omega)|t|^{-1/\mu}$ & $ x-\varepsilon\ln|t|$
  & $(\varphi^{\mu}\varphi')'=-\delta\varepsilon\varphi'-\delta\mu^{-1}\varphi$\\
7&$\langle Q_1+\varepsilon Q_4\rangle$ & $\varphi(\omega)e^{2\varepsilon\mu^{-1} t}$
       & $xe^{-\varepsilon t}$
  &$(\varphi^{\mu}\varphi')'=-\varepsilon\omega\varphi'+2\mu^{-1}\varepsilon\varphi$\\
8&$\langle Q_2+\alpha Q_4\rangle$ & $\varphi(\omega)|t|^{(2\alpha-1)/\mu}$ & $ x|t|^{-\alpha}$ &
   $(\varphi^{\mu}\varphi')'=\delta\mu^{-1}(2\alpha-1)\varphi-\delta\alpha\omega\varphi'$\\
\hline
\end{tabular}
\end{center}

For some of the reduced equations the general solutions are known.
For other ones we succeeded to find only particular solutions.
These solutions are following:
\begin{gather}\nonumber
u=|c_1x+c_0|^{\frac1{\mu+1}},
\quad
u=(c_0-\varepsilon\mu(x-\varepsilon t))^{\frac1\mu},
\quad
u=\left(-\frac\mu{\mu+2}\,\frac{(x+c_0)^2}{2t}+c_1|t|^{-\frac{\mu}{\mu+2}}\right)^{\frac1\mu},
\\ \nonumber
u=\left(-\frac\mu{\mu+2}\,\frac{(x+c_0)^2}{2t}+
c_1(x+c_0)^{\frac\mu{\mu+1}}|t|^{-\frac{\mu(2\mu+3)}{2(\mu+1)^2}}\right)^{\frac1\mu},
\\
u=\varphi(x-\varepsilon t)\quad \mbox{where}\
\int{\frac{\varphi^\mu}{c_1-\varepsilon\varphi}}d\varphi=x-\varepsilon t+c_0. \label{SolNDEumu}
\end{gather}

Equation~\eqref{solv5} admits multiplicative separation of variables. Namely, for all values of $\mu$
one can find solution in form
of the product of two functions of different arguments:
\begin{equation}\label{SolSepNDEumu}
u(t,x)=(b_1t+b_0)^{-1/\mu}f(x),
\end{equation}
where the function $f=f(x)$ is given implicitly
\[
\int\dfrac{f^\mu df}{\sqrt{c_1-\lambda f^{\mu+2}}}=\pm x+c_0,\qquad \lambda=\dfrac{2b_1}{\mu(\mu+2)}.
\]

These could be found also in~\cite{Ibragimov1994V1,Polyanin&Zaitsev2004,Barenblatt1952,Hill1989,
Samarskii&Galaktionov&Kurdyumov&Mikhailov1995,Dorodnitsyn1982,Zwillinger1989,
Pukhnachov1995,Aristov1999,Akhatov&Gazizov&Ibragimov1989,Zel'dovich&Kompaneets1950}.
The most studied cases are equations with $\mu=\pm1,-2,-4/3,-3/2$.
Below we adduce their exact solutions that are inequivalent to~\eqref{SolNDEumu}
with respect to the Lie symmetry transformations.

Equation with the singular value of the parameter~$\mu=-1$
called often the {\em fast diffusion equation},
is distinguished by the reduction procedure.
Lie invariant solutions of it will be adduced in separate section together with non-Lie solutions
obtained from invariance of the fast diffusion equation with respect to the nonclassical potential reduction operators.

Fujita--Storm equation~\eqref{eqFuSE} is linearizable and has been considered in a separate section.

Equation
\mathversion{bold}\begin{equation} \label{eqDEu-43}
u_t=\left(u^{-4/3}{u_x}\right)_x
\end{equation}\mathversion{normal}
admits the five-dimensional Lie symmetry algebra generated by
\[
Q_1=\p_t,\ Q_2=t\p_t-\mu^{-1}u\p_u,\ Q_3=\p_x,\ Q_4=x\p_x+\frac32u\p_u,\ Q_5=-x^2\p_x+3xu\p_u.
\]
An optimal system of one-dimensional subalgebras of this algebra is
\[
\langle Q_1+Q_4\rangle,\ \langle aQ_3+Q_4 \rangle,\ \langle Q_5 \rangle,\ \langle Q_2+Q_5 \rangle,\
\langle Q_3+Q_5 \rangle.
\]
One can easily construct the corresponding ansatzes and the reduced ODEs.
However, to the best of our knowledge all the found solutions of these equations are equivalent to~\eqref{SolNDEumu}
with particular value of parameter $\mu=-4/3$.
Solutions of~\eqref{eqDEu-43} invariant with respect to dilatation operators can be found also in~\cite{King1991}.
Besides the already adduced Lie invariant solutions, equation~\eqref{eqDEu-43} has functional
separated solution~\cite{Galaktionov1995,Rudykh&Semenov1998}
\[
u=(\varphi_4(t)x^4+\varphi_3(t)x^3+\varphi_2(t)x^2+\varphi_1(t)x+\varphi_0(t))^{-3/4},
\]
where the functions $\varphi_i=\varphi_i(t)$ are determined by the system of ordinary differential equations
\begin{gather*}
\varphi'_0=-\frac34\varphi_1^2+2\varphi_0\varphi_2,\quad \varphi'_1=-\varphi_1\varphi_2+6\varphi_0\varphi_3,\quad
\varphi'_2=-\varphi_2^2+\frac32\varphi_1\varphi_3+12\varphi_0\varphi_4,\\
\varphi_3'=-\varphi_2\varphi_3+6\varphi_1\varphi_4,\quad \varphi'_4=-\frac34\varphi_3^2+2\varphi_2\varphi_4.
\end{gather*}
The general form of exact solutions of~\eqref{eqDEu-43}
obtained from the known ones $u=f(t,x)$ with action of group transformations is
\[
\tilde u=\dfrac{\varepsilon_1^{-1}\varepsilon_2^2}{(\varepsilon_5x+1)^3}
f\left(\varepsilon_1^{-4/3} t+\varepsilon_3,\dfrac{\varepsilon_2^{-4/3} x}{\varepsilon_5x+1}+\varepsilon_4\right).
\]

Equation
\mathversion{bold}\[
u_t=\left(u^{-3/2}{u_x}\right)_x,
\]\mathversion{normal}%
admits also the functional separation of variables. The corresponding exact solution is
\[
u=(3c_1x^3+f_2(t)x^2+f_1(t)x+f_0(t))^{-2/3}.
\]
Here
\begin{gather*}\textstyle
f_2(t)=3\int\varphi(t)dt+3c_2,\quad f_1(t)=\dfrac1{c_1}(\int\varphi(t)dt+c_2)^2+\dfrac1{2c_1}\varphi(t),
\\ \textstyle
f_0(t)=\dfrac1{9c_1^2}(\int\varphi(t)dt+c_2)^3+\dfrac1{6c_1^2}\varphi(t)(\int\varphi(t)dt+c_2)
+\dfrac1{36c_1^2}\varphi'(t),
\end{gather*}
where the function $\varphi=\varphi(t)$ is defined implicitly by $\int(c_3-8\varphi^3)^{-1/2}d\varphi=\pm t+c_4$ .

T.K.~Amerov~\cite{Amerov1990} and J.R.~King~\cite{King1992} suggested to look for solutions of the equation
\mathversion{bold}\[
u_t=(u^{-1/2}u_x)_x
\]\mathversion{normal}%
in the form $u=(\varphi^1(x)t+\varphi^0(x))^2$ where the functions $\varphi^1(x)$ and $\varphi^0(x)$
satisfy the system of ODEs
$\varphi^1_{xx}=(\varphi^1)^2,$ $\varphi^0_{xx}=\varphi^0\varphi^1.$
A particular solution of this system is
\[
\varphi^1=\frac6{x^2} , \qquad  \varphi^0=\frac{c_1}{x^2}+\frac{c_2}{x^3}.
\]

\section{Porous medium equation \mathversion{bold}$u_t=(uu_x)_x$}\label{SectionOnSolutionOfPorMedEq}

Another important subclass of diffusion equations is a special case of equation~\eqref{solv5} with $\mu=1$
\mathversion{bold}\begin{equation}\label{eqPorMedEq}
u_t=(uu_x)_x,
\end{equation}\mathversion{normal}
called also {\em porous medium equation}.
It first exact solution has been obtained by Boussinesq~\cite{Boussinesq1904}.
He was looking for a solution in a separated form $u(t,x)=X(x)T(t)$ satisfying conditions
\[
u(t,0)=0,\quad u_x|_{x=L}=0.
\]
Thus constructed solution reads as
\[
u=\frac{H_0F(\xi)}{1+(3b^2H_0/2L^2)t},
\]
where $H_0$ is a constant, $\xi=x/L$ and the function $F=F(\xi)$ is defined implicitly
\[
\xi=\frac1b\int_0^F\frac{\lambda d\lambda}{\sqrt{1-\lambda^3}}, \quad
b=\int_0^1\frac{\lambda d\lambda}{\sqrt{1-\lambda^3}}=\frac13B\Big(\frac23,\frac12\Big).
\]
The next exact solution of the porous medium equation~\eqref{eqPorMedEq}
was found much later by Barenblatt~\cite{Barenblatt1952} and written in the present form by
Sokolov~\cite{Sokolov1956} (the instant source solution):
\[
u=\frac1{6t}((9t)^{2/3}-x^2),\quad 0\le x\le (9t)^{1/3}=l.
\]
It is easy to become convinced of the fact that both Boussinesq and Barenblatt solutions
correspond to the Lie symmetry of the porous medium equation~\eqref{eqPorMedEq}~\cite{Zhdanov&Lahno1998}.
Indeed, the Boussinesq solution is a particular case of the ansatz
\[
u=(1+\alpha t)^{-1}\phi(x),\quad \alpha=\frac{3b^2H_0}{2L^2},
\]
that is invariant with respect to the one-parameter Lie symmetry group generated by
$Q=(1+\alpha t)\p_t-u\p_u$.
Barenblatt solution is invariant with respect to the one-parameter Lie symmetry group generated by
$Q=3t\p_t+x\p_x-u\p_u$.

Besides~\eqref{SolNDEumu}, its exact solutions in parametric form are known~\cite{Polyanin&Zaitsev2004}:
\begin{gather*}
x=(6t+c_1)\xi+c_2\xi^2+c_3,\quad u=-(6t+c_1)\xi^2-2c_2\xi^3,\\
x=tf(\omega)+g(\omega), \quad u=tf'(\omega)+g'(\omega),
\end{gather*}
where the functions $f=f(\omega)$ and $g=g(\omega)$are determined by the system of ODEs:
\[
(f')^2-ff''=f''',\quad f'g'-fg''=g'''.
\]
It is obvious, that the second equation has two linearly independent particular solutions $g=1$ and $g=f$.
The general solution of these equations can be represented in form
\[\textstyle
g=c_1+c_2f+c_3(f\int\psi d\omega-\int f\psi d\xi),\quad f=f(\omega),\quad\psi=\dfrac1{(f')^2}e^{-\int fd\omega}.
\]
It is not difficult to verify, that it has the following particular solutions
\[
f=\dfrac{6}{\omega+c_1}\quad\mbox{and}\quad f=c_1e^{c_2\omega}.
\]
One can see, that the first solution leads to the previously given implicit solution.

\section{Fast diffusion equation \mathversion{bold}$u_t=(u^{-1}u_x)_x$\mathversion{normal}}
\label{SectionOnSolutionOfFastDifEq}

All invariant solutions of fast diffusion equation
\mathversion{bold}\begin{equation} \label{eqFDE}
u_t=\left(u^{-1}{u_x}\right)_x,
\end{equation}\mathversion{normal} which were earlier constructed in closed forms with the classical Lie method,
were collected e.g. in~\cite{Polyanin&Zaitsev2004,Popovych&Ivanova2005PETs,Popovych&Vaneeva&Ivanova2005,
Popovych&Ivanova2004NVCDCEs}.
A complete list of $G_1$-inequivalent solutions of such type is exhausted by the following ones:
\begin{gather}\label{Liesolutions.for.u-1}
\begin{split}&
1)\ u=\dfrac{1}{1+\varepsilon e^{x+t}},\qquad
2)\ u=e^{x},\qquad
3)\ u=\dfrac{1}{x-t+\mu te^{-x/t}},
\\[1ex]&
4)\ u=\dfrac{2t}{x^2+\varepsilon t^2},\qquad
5)\ u=\dfrac{2t}{\cos^2x},\qquad
6)\ u=\dfrac{-2t}{\cosh^2x},\qquad
7)\ u=\dfrac{2t}{\sinh^2x}.
\end{split}\end{gather}
The below arrows denote the possible transformations of solutions~\eqref{Liesolutions.for.u-1}
to each other by means of the potential hodograph transformation~\eqref{pothodograph}
up to translations with respect to~$x$~\cite{Popovych&Ivanova2005PETs}:
\[
\begin{split}&
\mbox{\Large$\circlearrowright$}\;1)_{\varepsilon=0}\,; \quad
1)_{\varepsilon=1}\longleftrightarrow 1)_{\varepsilon=-1,\;x+t<0}\,; \quad
\mbox{\Large$\circlearrowright$}\;1)_{\varepsilon=-1,\;x+t>0}\,; \quad
2)\longleftrightarrow 3)_{\mu=0,\;x>t}\,; \\&
\mbox{\Large$\circlearrowright$}\;4)_{\varepsilon=0}\,; \quad
5)\longleftrightarrow 4)_{\varepsilon=4}\,; \quad
6)\longleftrightarrow 4)_{\varepsilon=-4,\;|x|<2|t|}\,; \quad
7)\longleftrightarrow 4)_{\varepsilon=-4,\;|x|>2|t|}\,.
\end{split}
\]
The sixth connection can be found also in~\cite{Fushchych&Serov&Amerov1982,Pukhnachov1996}.
If $\mu\ne0$ solution 3) from list~\eqref{Liesolutions.for.u-1} is mapped by~\eqref{pothodograph}
to the solution
\[
8)\ u=t\vartheta(\omega)-t+\mu te^{-\vartheta(\omega)},\qquad \omega=x-\ln|t|,
\]
which is invariant with respect to the algebra~$\langle t\p_t+\p_x+u\p_u\rangle$.
Here $\vartheta$ is the function determined implicitly by the formula
$
\int (\vartheta-1+\mu e^{-\vartheta})^{-1}d\vartheta=\omega.
$

Some non-Lie exact solutions of~\eqref{eqFDE} were obtained in~\cite{Rosenau1995,Qu1999,Gandarias2001}.
Thus, P.\,Rosenau~\cite{Rosenau1995} found that potential equation $v_t=v_x{}^{-1}v_{xx}$ corresponding
to~\eqref{eqFDE} admits,
in addition to the usual variable separation $v=T(t)X(x)$, the additive one $v=Y(x+\lambda t)+Z(x-\lambda t)$
which is a potential additive variable separation for the fast diffusion equation~\eqref{eqFDE}.
(The classical multiplicative separation of variables is given by~\eqref{SolSepNDEumu} with $\mu=-1$.)
To construct nonclassical solutions of~\eqref{eqFDE},
C.~Qu~\cite{Qu1999} made use of generalized conditional symmetry method, looking for the conditional symmetry
operators in the special form $Q=(u_{xx}+H(u){u_x}^2+F(u)u_x+G(u))\p_u$.
M.L.\,Gandarias~\cite{Gandarias2001} investigated some families of usual and potential nonclassical symmetries
of~\eqref{solv5}. In particular, using an ansatz for the coefficient~$\eta$, she found
non-trivial reduction operators in the so-called ``no-go''
case when the coefficient of~$\p_t$ vanishes,
i.e. operators can be reduced to the form~$Q=\p_x+\eta(t,x,u)\p_u$.
These solutions and the ones similar to them were represented uniformly over the complex field
as compositions of two simple waves which move with the same ``velocities''
in opposite directions in~\cite{Popovych&Vaneeva&Ivanova2005}.
Using such representation the following solutions of fast diffusion equation~\eqref{eqFDE}
were obtained~\cite{Popovych&Vaneeva&Ivanova2005}:
\[\begin{split}&
1')\ u=\cot(x-t)-\cot(x+t)=\dfrac{2\sin 2t}{\cos 2t-\cos 2x},
\\&
2')\ u=\coth(x-t)-\coth(x+t)=\dfrac{2\sinh 2t}{\cosh2x-\cosh 2t},\\[1ex]&
3')\ u=\coth(x-t)-\tanh(x+t)=\dfrac{2\cosh 2t}{\sinh 2x-\sinh 2t},\\[1ex]&
4')\ u=\tanh(x-t)-\tanh(x+t)=-\dfrac{2\sinh 2t}{\cosh 2x+\cosh 2t},\quad\\[1ex]&
5')\ u=\cot(ix+t)-\cot(ix-t)=\dfrac{2\sin 2t}{\cosh 2x-\cos 2t},\\[1ex]&
6')\ u=i\cot(x+it)-i\cot(x-it)=\dfrac{2\sinh 2t}{\cosh 2t-\cos 2x}.
\end{split}\]

Transformation~(\ref{pothodograph}) acts on the set of solutions $1')$--$6')$
in the following way~\cite{Popovych&Vaneeva&Ivanova2005}:
\[
\begin{split}&
1')_{\cos 2t<\cos 2x} \longleftrightarrow 5')|_{t\to t+\pi/2,\,x\to x/2,\,v\to 2v}; \quad
1')_{\cos 2t>\cos 2x} \longleftrightarrow 5')|_{x\to x/2,\,v\to 2v-\pi}\,;
 \quad
\\&
2')_{|x|<|t|}\longleftrightarrow 4')|_{x\to x/2,\,v\to 2v}; \quad
\mbox{\Large$\circlearrowright$}\;2')_{|x|>|t|}|_{x\to x/2,\,v\to 2v}\,;\quad
\\&
\mbox{\Large$\circlearrowright$}\;3')_{x<t}|_{x\to x/2,\,v\to 2v}; \quad
3')_{x>t} \longleftrightarrow 3')_{x>t}|_{x\to -x/2,\,v\to -2v}; \quad
\mbox{\Large$\circlearrowright$}\;6')|_{x\to x/2,\,v\to 2v}
.
\end{split}
\]
These actions can be interpreted in terms of actions of transformation~(\ref{pothodograph}) on
the nonclassical symmetry operators which correspond to solutions $1')$--$6')$.

In~\cite{Rosenau1995} P.~Rosenau considered additive separation of variables
for the potential fast diffusion equation~\eqref{eqFDE} and constructed solution $4')$.
Using the generalized conditional symmetry method, C.~Qu~\cite{Qu1999} found
solutions which can be written in forms $1')$ and $6')$.
After rectifying computations in two cases from~\cite{Qu1999}, one
can find also solutions $2')$ and $5')$.
Solutions $1')$, $3')$ and $4')$ were obtained in \cite{Gandarias2001}.
The remaining solutions from the above list were found in~\cite{Popovych&Vaneeva&Ivanova2005}.

One of techniques which can be applied for finding the above solutions is reduction by conditional symmetry operators
of the form $Q=\p_x+(\eta^1(t,x)u+\eta^2(t,x))u\p_u$ (see~\cite{Gandarias2001} for details).
All reductions performed with reduction operators of such type
result in solutions which are equivalent to the listed Lie solutions 1)--7) or solutions $1')$--$6')$.

\pagebreak

\section{Nonlinear diffusion equations. Other nonlinearities}\label{SectionOnSolutionsOfNDEMisc}

Known exact solutions of the reduced equation~\ref{TableReductionForNDE}.4 corresponding to
$\langle Q_3=2t\p_t+x\p_x\rangle$
(which are self-similar solutions of~\eqref{eqNDE}) are adduced in
Table~\ref{TableSelfSimSolForNDE}~\cite{Polyanin&Zaitsev2004}.

{\begin{center}\refstepcounter{tabul}\label{TableSelfSimSolForNDE}
Table~\thetabul. Self-similar solutions for nonlinear diffusion equations~\eqref{eqNDE}, $\omega=x/\sqrt{t}$
\\[1.5ex] \footnotesize
\setcounter{tbn}{0}
\renewcommand{\arraystretch}{1.2}
\begin{tabular}{|l|c|c|l|}
\hline \vspacebefore
N&$A(u)$& Solution $\omega=\omega(u)$ &\hfill {Conditions\hfill} \\
\hline
\vspacebeforemore\refstepcounter{tbn}\thetbn&$\dfrac n2u^n-\dfrac n{2(n+1)}u^{2n} $ & $1-u^n$ & $n>0$\\[1ex]
\hline
\vspacebeforemore\refstepcounter{tbn}\thetbn&$\dfrac n{2(n+1)}((1-u)^{n-1}-(1-u)^{2n}) $ & $(1-u)^n$ & $n>0$\\[1ex]
\hline
\vspacebeforemore\refstepcounter{tbn}\thetbn&$\dfrac n{2(1-n)}u^{-2n}-\dfrac n2u^{-n} $ & $u^{-n}-1$ & $0<n<1$\\[1ex]
\hline
\vspacebeforemore\refstepcounter{tbn}\thetbn&$\dfrac12\sin^2\dfrac{\pi u}2 $ & $\cos\dfrac{\pi u}2$ & \\[1ex]
\hline
\vspacebeforemore\refstepcounter{tbn}\thetbn&$\dfrac18\sin{\pi u}\,(\pi u+\sin\pi u) $ & $\cos^2\dfrac{\pi u}2$ & \\[1ex]
\hline
\vspacebeforemore\refstepcounter{tbn}\thetbn&$\dfrac1{16}\sin^2{\pi u}\,(5+\cos\pi u) $ & $\cos^3\dfrac{\pi u}2$ &\\[1ex]
\hline
\vspacebeforemore\refstepcounter{tbn}\thetbn
    &$\dfrac12\cos\dfrac{\pi u}2\,\left(\cos\dfrac{\pi u}2+\dfrac{\pi u}2-1\right)$ & $1-\sin\dfrac{\pi u}2$ & \\[1ex]
\hline
\vspacebeforemore\refstepcounter{tbn}\thetbn&$\dfrac{u\arccos u+1}{2\sqrt{1-u^2}}-\dfrac12$ & $\arccos u$ & \\[1ex]
\hline
\vspacebeforemore\refstepcounter{tbn}\thetbn&$\dfrac{\pi-2(1-u)\arcsin(1-u)}{4\sqrt{2u-u^3}}-\dfrac12 $
   & $\arcsin(1-u)$ & \\[1ex]
\hline
\vspacebeforemore\refstepcounter{tbn}\thetbn&$\dfrac{u\arcsin u}{4\sqrt{1-u^2}}+\dfrac14u^2$ & $\sqrt{1-u^2}$ & \\[1ex]
\hline
\vspacebeforemore\refstepcounter{tbn}\thetbn&$\dfrac12(1-\ln u)$ & $-\ln u$ & \\[1ex]
\hline
%\vspacebefore\refstepcounter{tbn}\thetbn&$\sinh^2 u$ & $\pm\sqrt2\cosh u$ & \\
%\hline
%\vspacebefore\refstepcounter{tbn}\thetbn&$\cosh^2 u$ & $\pm\sqrt2\sinh u$ & \\
%\hline
\end{tabular}
\end{center}}

Let us give some more examples of travelling wave
solutions~\cite{Samarskii&Galaktionov&Kurdyumov&Mikhailov1995,Polyanin&Zaitsev2004,Ibragimov1994V1}:
\begin{gather*}
\mbox{\mathversion{bold}$u_t=(\sinh^2 u u_x)_x$,\mathversion{normal}}\quad
   u=\arccosh^2\dfrac{\pm x+c_1}{\sqrt{c_2-2t}},\\
\mbox{\mathversion{bold}$u_t=(\cosh^2 u u_x)_x$},\quad u=\arcsinh^2\dfrac{\pm x+c_1}{\sqrt{c_2-2t}},\\
\mbox{\mathversion{bold}$u_t=((u^{2\mu}+bu^\mu)u_x)_x$,\mathversion{normal}}\quad
u=\left(\sqrt{\dfrac{c_1}{a(c_1+1)(c_2-t)}}( x+c_1)-\dfrac{2bc_1}{a(c_1+1)}\right)^{1/\mu}.
\\
\mbox{\mathversion{bold}$u_t=((e^{2u}+bue^u)u_x)_x$,\mathversion{normal}}\quad
u=\ln\left(\dfrac{\pm x+c_1}{\sqrt{c_2-2t}}-b\right).
\\
\mbox{\mathversion{bold}$u_t=(ue^uu_x)_x$,\mathversion{normal}}\quad
u=\ln(c_1x+c_1^2t+c_0).
\end{gather*}

For equation with logarithmical nonlinearity
\mathversion{bold}\[
u_t=(\ln u u_x)_x
\]\mathversion{normal}
travelling wave and self-similar solutions are known:
\[
u=\exp\left(\pm\sqrt{2c_1x+2c_1^2t+c_2}\right),\quad u=\exp\left(\dfrac{\pm x+c_1}{c_2-2t}-1\right).
\]

A number of exact solution for equations of class~\eqref{eqNDE} were constructed with nonlocal (quasilocal or potential)
symmetries~\cite{Akhatov&Gazizov&Ibragimov1989,Bluman&Kumei1989,Bluman&Reid&Kumei1988,Sophocleous1996,
Sophocleous2000,Sophocleous2003}.

Thus, e.g., reductions with respect to the optimal system of subalgebras of Lie algebra
of potential/quasilocal symmetries of
equation
\mathversion{bold}\[
u_t=((1+u^2)^{-1}u_x))_x
\]\mathversion{normal}
give rise to exact solutions of form
\begin{gather*}
u=ce^{t-x}(1-c^2e^{2(t-x)})^{-1/2},\quad u=-x(c-2t-x^2)^{-1/2},\\
u=\tan(\varphi(\omega)+\arctan(\lambda(\omega))+\varepsilon t),\quad\mbox{where}\quad
\omega=x^2+v^2,\quad v=\tan(\varphi(\omega)+\varepsilon t),\\
u=x\tan(\varphi(\omega)+\arctan(\lambda(\omega))+\varepsilon t),\quad\mbox{where}\quad \omega=\dfrac{x^2+v^2}t,\quad
               v=x\tan\left(\varphi(\omega)+\dfrac\alpha2\varepsilon t\right).
\end{gather*}
Here $\varphi(\omega)$ and $\lambda(\omega)$ are arbitrary solutions of the system
$\varphi'=\omega^{-1}\lambda/2$, $\lambda'=(1+\lambda^2)(\varepsilon-\omega^{-1}\lambda)/2$.
The list of known exact solutions of this equation (called often the Fujita's equation) involves also the following
ones~\cite{Ibragimov1994V1,Polyanin&Zaitsev2004,Akhatov&Gazizov&Ibragimov1989,Doyle&Vassiliou1998}:
\begin{gather*}
u=\tan(c_1x+c_2),\quad u=\pm x(c_1-2t-x^2)^{-1/2},\\
u=\dfrac{\pm e^{t-x}}{\sqrt{1-e^{2(t-x)}}},\quad
\varepsilon(x+\varepsilon t)+c_2=\dfrac{1}{c_1^2+1}\left(\ln\dfrac{|u+c_1|}{\sqrt{u^2+1}}+c_1\arctan u\right),\\
u=\dfrac{v}{\sqrt{1-v^2}},\quad \mbox{where}\quad
v=\dfrac{c_1e^{\lambda x}+c_2e^{-\lambda x}}{\sqrt{4c_1c_2+c_3e^{-2\lambda^2t}}} \quad\mbox{or}\\
v=\dfrac{c_1\sin\lambda x+c_2\cos\lambda x}{\sqrt{c_1^2+c_2^2+c_3e^{2\lambda^2t}}},\\
u=\frac{\sinh x}{\sqrt{-\cosh^2x-e^{-2t}}},\quad
u=\frac{\pm\cosh x}{\sqrt{-\sinh^2x+e^{-2t}}},\quad
u=\frac{\sin x}{\sqrt{\cos^2x\pm e^{2t}}}.
\end{gather*}

We adduce also some exact solutions of another equation with Fujita's type nonlinearity
\mathversion{bold}\[
u_t=((1-u^2)^{-1}u_x))_x,
\]\mathversion{normal}
namely~\cite{Doyle&Vassiliou1998}:
\begin{gather*}
u=c,\quad u=\tanh x,\quad u=\frac{x}{\sqrt{x^2+2t}}\ (t>0), \quad u=\frac{\pm e^{t-x}}{\sqrt{1+e^{{2(x-t)}}}},\\
u=\frac{\sinh x}{\sqrt{\cosh^2x+e^{-2t}}},\quad u=\frac{\sinh x}{\sqrt{\cosh^2x-e^{-2t}}}\ (t>0),
\quad u=\frac{\pm\cosh x}{\sqrt{\sinh^2x+e^{-2t}}}\ (t<0),\\
u=\frac{\sin x}{\sqrt{-\cos^2x+e^{2t}}}
\end{gather*}
The third similarity solution converges (pointwise) to a step function as $t\to0^+$, and to zero as $t\to\infty$.
The fourth solutions are bounded travelling waves. The fifth and sixth solutions converge to the time independent second solution as $t\to\infty$.
The sixth solution converge to a step function as $t\to0^+$.
The seventh solution converges to the values $\pm1$ as $t\to0-$. The last solution converges to a square wave as $t\to0+$, and to zero as $t\to\infty$.

One more example of solution obtained with application of potential symmetry is
$u=\tan(\varphi(\omega)+\arctan(2\omega\varphi')-\lambda^{-1}\ln t)$
for the equation
\mathversion{bold}\[
u_t=((1+u^2)^{-1}e^{\lambda\arctan u}u_x))_x.
\]\mathversion{normal}
Here
\[
\omega=x^2+v^2,\quad \varphi=\dfrac1\lambda\ln\dfrac{c-\omega}2-\arctan\psi(\omega),\quad
\dfrac{\psi'}{1+\psi^2}+\dfrac\psi{2\omega}+\dfrac2{\lambda(c-\omega)}=0.
\]

All the potential symmetries of equations from class~\eqref{eqCCDCE} can be obtained
from Lie symmetries of~\eqref{eqCCDCE} by means of prolongation to the potential variable $v$
and application of potential equivalence transformations~\cite{Popovych&Ivanova2005PETs}
\[
\tilde t=t,\quad
\tilde x=x+\varepsilon v,\quad
\tilde u=\dfrac u{1+\varepsilon u},\quad
\tilde v=v,\quad
\tilde A=(1+\varepsilon u)^2A
\]
and hodograph transformation%~\eqref{pothodograph}
\begin{gather}\label{pothodograph}
\tilde t=t,\quad
\tilde x=v,\quad
\tilde u=u^{-1},\quad
\tilde v=x,\quad
\tilde A=u^2A,
\end{gather}
where $v_x=u$, $v_t=Au_x$.

Therefore, these transformations can be used for obtaining potentially invariant exact solutions from the Lie ones.
The complete list of nonlinear constant coefficient diffusion and diffusion--convection equations
having potential symmetries together with the transformations mapping them to
the equations with power and exponential nonlinearities can be found in~\cite{Popovych&Ivanova2005PETs}.

In~\cite{Tsyfra&Messina&Napoli&Tretynyk2004} non-point nonclassical symmetry operators are used
to obtain exact solutions of evolution equations.
In particular, it is shown that equation
\mathversion{bold}\[
u_t=(u^{-2}e^{1/u}u_x))_x
\]\mathversion{normal}
admits an exact solution in implicit form $u=(z^2+c)/(2t)$, where
\begin{gather*}
z=\dfrac z{\ln\frac{z^2+c}{2e^2t}+\frac{2\sqrt c}z\arctan\frac z{\sqrt c}+\frac{c_1}z},\quad\mbox{if}\quad
c>0,\quad\mbox{and}\\
z=\dfrac z{\ln\frac{z^2+c}{2e^2t}+\frac{2\sqrt {-c}}z\ln\frac{z-\sqrt{-c}}{z+\sqrt{-c}}+\frac{c_1}z},\quad
\mbox{if}\quad c<0.
\end{gather*}

A number of authors considered additive separated solutions of diffusion equations~\eqref{eqNDE}, i.e., solutions of form
\[
u(t,x)=\varphi(t)+\psi(x).
\]
Usually such solutions are Lie invariant. They were adduced in previous sections.
So, here we adduce only list of equations admitting such kind of separation of variables.
Namely, a diffusion equation~\eqref{eqNDE} admits separation of variables if and only if it is $G^{\sim}$-equivalent to equation
with the diffusion coefficient being of the following functions~\cite{Doyle&Vassiliou1998}:
\begin{gather*}
A=|u|^\mu, \quad A=e^u, \quad A=(u^2\pm1)^{-1},\quad\\
A=z(u)e^{z(u)},\quad u=\int_1^zs^{-3/2}e^{-s/2}ds,\quad z>0,\\
A=e^{\sigma z(u)}\cosh z(u), \quad \sigma\ne\pm1,\quad u=\int_0^z \cosh^{-3/2}se^{-\sigma s/2}ds,\quad -\infty<z<\infty\\
A=e^{\sigma z(u)}\sinh z(u), \quad \sigma\ne\pm1,\quad u=\int_1^z \sinh^{-3/2}se^{-\sigma s/2}ds,\quad z>0,\\
A=e^{\sigma z(u)}\cos z(u), \quad u=\int_0^z \cos^{-3/2}se^{-\sigma s/2}ds,\quad -\pi/2<z<\pi/2.
\end{gather*}

\section{Constant coefficient diffusion--convection equations}\label{SectionOnSolutionsOfNCCDCE}

Lie symmetries of the constant coefficient diffusion--convection equation
\mathversion{bold}\begin{equation}\label{eqCCDCE}
u_t=(A(u)u_x)_x+B(u)u_x,
\end{equation}\mathversion{normal}%
$B\ne0$ and corresponding Lie reductions were considered by a number of authors,
see for example,~\cite{Katkov1965,Katkov1968,Oron&Rosenau1986,Edwards1994,Yung&Verburg&Baveye1994}.
However the complete group classification of class~\eqref{eqCCDCE}
 was presented only recently in~\cite{Popovych&Ivanova2004NVCDCEs}.

Any equivalence transformation of class~\eqref{eqCCDCE} has the form:
\begin{gather}\label{EquivGroupNCCDCE}
\begin{split}
&\tilde t=t\varepsilon_4^2\varepsilon_5+\varepsilon_1, \quad
 \tilde x=x\varepsilon_4+\varepsilon_7 t+\varepsilon_2, \quad
 \tilde u=u\varepsilon_6+\varepsilon_3, \\
&\tilde A=A\varepsilon_5^{-1}, \quad
 \tilde B=B\varepsilon_4^{-1}\varepsilon_5^{-1}-\varepsilon_7,
\end{split}
\end{gather}
where $\varepsilon_1,\ldots,\varepsilon_7$ are arbitrary constants,
$\varepsilon_4\varepsilon_5\varepsilon_6\ne0$. Note, that all the equations with $B(u)=\const$
are reducible to diffusion equation~\eqref{eqNDE}. Besides such `trivial' cases,
Burgers equation and Fokas--Yortsos equation,
only few Lie invariant solutions for equations with non-zero convectivity are found.
Thus, e.g., in~\cite{Yung&Verburg&Baveye1994}
scale-invariant solution
\[
xu^{-\mu/2}-(1+2/\mu)\ln(xu^{-\mu/2}-1-2/\mu)=c_1t+c_0
\]
in implicit form is found for the equation with power nonlinearities
\mathversion{bold}\[
u_t=(u^\mu u_x)_x+u^{\mu/2}u_x.
\]\mathversion{normal}
was found.

M.~Edwards~\cite{Edwards1994} investigated Lie symmetries of~\eqref{eqCCDCE}
and constructed optimal subalgebras of the symmetry algebras
for some of equations from the class. Here we supplement her results and adduce the complete list of Lie reductions
of equations from class~\eqref{eqCCDCE}.
(The linearizable Fokas--Yortsos and Burgers equations have been considered separately
in Section~\ref{SectionOnSolutionOfLinearizEq}.)

{\begin{center}\refstepcounter{tabul}\label{TableReductionForCCEeueu}
Table~\thetabul. Reduced ODEs for $u_t=(e^{\mu u}u_x)_x+e^uu_x $ ($\mu\ne0$)
\\[1.5ex] \footnotesize
\setcounter{tbn}{0}
\renewcommand{\arraystretch}{1.2}
\begin{tabular}{|l|c|c|l|}
\hline \vspacebefore
Subalgebra& Ansatz $u=$& $\omega$ &\hfill {Reduced ODE\hfill} \\
\hline
$\langle \p_x\rangle$ & $\varphi(\omega)$ & $t$ & $\varphi'=0$\\
$\langle \p_t\rangle$ & $\varphi(\omega)$ & $x$ & $ (e^{\mu \varphi}\varphi')'+e^\varphi\varphi'=0$\\
$\langle \p_t+\varepsilon\p_x\rangle$ & $\varphi(\omega)$
  & $x-\varepsilon t$ & $-\varepsilon\varphi'=(e^{\mu \varphi}\varphi')'+e^\varphi\varphi'$\\
$\langle (\mu t-2t+\varepsilon)\p_t$ & $\varphi(\omega)+$ & $x((\mu-2)t+\varepsilon)^{\frac{1-\mu}{\mu-2}} $
   & $\varphi'(1-\mu)\omega+1=(e^{\mu \varphi}\varphi')'+e^\varphi\varphi'$\\
    $+(\mu-1)x\p_x+\p_u\rangle$ ($\mu\ne2$)& $+\dfrac{\ln((\mu-2)t+\varepsilon)}{\mu-2}$&&\\[1ex]
$\langle \varepsilon\p_t+x\p_x+\p_u\rangle$ ($\mu=2$) & $\varphi(\omega)+\dfrac t{\varepsilon}$
  & $xe^{-t/\varepsilon} $ & $\dfrac1\varepsilon-\varphi'\omega=(e^{2 \varphi}\varphi')'+e^\varphi\varphi'$\\
\hline
\end{tabular}
\end{center}}
If $u=f(t,x)$ is an exact solution of equation $u_t=(e^{\mu u}u_x)_x+e^uu_x $, then
\[
u=f\big(e^{(\mu-2)\varepsilon_1}t+\varepsilon_3,e^{(\mu-1)\varepsilon_1+\varepsilon_2}x
+\varepsilon_4\big)+\varepsilon_1+\varepsilon_2
\]
is also solution of the same equation.
%
%Some of these equations can be integrated

\bigskip

{\begin{center}\refstepcounter{tabul}\label{TableReductionForCCEeuu}
Table~\thetabul. Reduced ODEs for $u_t=(e^{u}u_x)_x+uu_x $.
\\[1.5ex] \footnotesize
\setcounter{tbn}{0}
\renewcommand{\arraystretch}{1.2}
\begin{tabular}{|l|c|c|l|}
\hline \vspacebefore
Subalgebra& Ansatz $u=$& $\omega$ &\hfill {Reduced ODE\hfill} \\
\hline
$\langle \p_x\rangle$ & $\varphi(\omega)$ & $t$ & $\varphi'=0$\\
$\langle \p_t\rangle$ & $\varphi(\omega)$ & $x$ & $ (e^\varphi\varphi')'+\varphi\varphi'=0$\\
$\langle \p_t+\varepsilon\p_x\rangle$ & $\varphi(\omega)$ & $x-\varepsilon t$
   & $-\varepsilon\varphi'=(e^\varphi\varphi')'+\varphi\varphi'$\\
$\langle (t+\varepsilon)\p_t+(x-t)\p_x+\p_u\rangle$ & $\varphi(\omega)+\ln|t+\varepsilon| $
& $\dfrac{x+\varepsilon}{t+\varepsilon}+\ln|t+\varepsilon|$
   & $1+\varphi'(1-\omega)=(e^\varphi\varphi')'+\varphi\varphi'$\\
\hline
\end{tabular}
\end{center}}
%Case~\ref{TableReductionForCCEeuu}.3 leads to the known travelling wave solution
%\[
%2\int\frac{e^udu}{-u^2+2u+c_1}=x-\varepsilon t
%\]
If $u=f(t,x)$ is an exact solution of equation $u_t=(e^{u}u_x)_x+uu_x$, then
\[
u=f\big(e^{\varepsilon_1}t+\varepsilon_2,e^{\varepsilon_1}(x-\varepsilon_1t+\varepsilon_3)\big)+\varepsilon_1
\]
is also solution of the same equation.

\bigskip

{\begin{center}\refstepcounter{tabul}\label{TableReductionForCCumuunu}
Table~\thetabul. Reduced ODEs for $u_t=(u^{\mu}u_x)_x+u^\nu u_x $.
\\[1.5ex] \footnotesize
\setcounter{tbn}{0}
\renewcommand{\arraystretch}{1.2}
\begin{tabular}{|l|c|c|l|}
\hline \vspacebefore
Subalgebra& Ansatz $u=$& $\omega$ &\hfill {Reduced ODE\hfill} \\
\hline
$\langle \p_x\rangle$ & $\varphi(\omega)$ & $t$ & $\varphi'=0$\\
$\langle \p_t\rangle$ & $\varphi(\omega)$ & $x$ & $ (\varphi^{\mu}\varphi)'+\varphi^\nu \varphi'=0$\\
$\langle \p_t+\varepsilon\p_x\rangle$ & $\varphi(\omega)$ & $x-\varepsilon t$
  & $-\varepsilon\varphi'=(\varphi^{\mu}\varphi')'+\varphi^\nu \varphi'$\\
$\langle (\mu t-2\nu t+\varepsilon)\p_t$
 & $\varphi(\omega)((\mu-2\nu)t+\varepsilon)^{\frac{1}{\mu-2\nu}}$
 & $x((\mu-2\nu)t+\varepsilon)^{\frac{\nu-\mu}{\mu-2\nu}} $
  & $\varphi+\varphi'\omega= (\varphi^{\mu}\varphi')'+\varphi^\nu \varphi'$\\
    $+(\mu-\nu)x\p_x+u\p_u\rangle$, ($\mu\ne2\nu$) &&&\\
$\langle \varepsilon\p_t+\nu x\p_x+u\p_u\rangle$, ($\mu=2\nu$)
 & $\varphi(\omega)e^{t/\varepsilon}$
 & $xe^{-\nu t/\varepsilon} $ & $\varphi-\varphi'\omega= (\varphi^{2\nu}\varphi')'+\varphi^\nu \varphi'$\\
\hline
\end{tabular}
\end{center}}

\bigskip

If $u=f(t,x)$ is an exact solution of equation $u_t=(u^{\mu}u_x)_x+u^\nu u_x $, then
\[
u=e^{\varepsilon_1+\varepsilon_2}f\big(e^{(\mu-2\nu)\varepsilon_1}t+\varepsilon_3,
  e^{(\mu-\nu)\varepsilon_1+\nu\varepsilon_2}x+\varepsilon_4\big)
\]
is also solution of the same equation.

For equations with $\nu=\mu$ the generalized travelling wave solution is known:
\[
u=\left(\dfrac{c_2-x}{t+c_1}+\dfrac{\ln|t+c_1|}{\mu(t+c_1)}\right)^{1/\mu}.
\]

If $\mu=2$, $\nu=1$ then the generalized travelling wave solution in implicit form is the following:
\[
2\int\dfrac{u^2du}{-u^2-2\varepsilon u+c_1}=x-\varepsilon t+c_2.
\]
For such values $\mu$ and $\nu$ the degenerate solution linear in $x$ has the form $u=\tau(t)(x+c_1)$,
where function $\tau$ is given in implicit form
\[
-\dfrac1\tau+2\ln\left|\dfrac{2\tau+1}\tau\right|=t+c_2.
\]

{\begin{center}\refstepcounter{tabul}\label{TableReductionForCCumulnu}
Table~\thetabul. Reduced ODEs for $u_t=(u^{\mu}u_x)_x+\ln u u_x $.
\\[1.5ex] \footnotesize
\setcounter{tbn}{0}
\renewcommand{\arraystretch}{1.2}
\begin{tabular}{|l|c|c|l|}
\hline \vspacebefore
Subalgebra& Ansatz $u=$& $\omega$ &\hfill {Reduced ODE\hfill} \\
\hline
$\langle \p_x\rangle$ & $\varphi(\omega)$ & $t$ & $\varphi'=0$\\
$\langle \p_t\rangle$ & $\varphi(\omega)$ & $x$ & $ (\varphi^{\mu}\varphi)'+\ln\varphi\varphi'=0$\\
$\langle \p_x+\varepsilon\p_t\rangle$ & $\varphi(\omega)$ & $x-\varepsilon t$
   & $-\varepsilon\varphi'=(\varphi^{\mu}\varphi)'+\ln\varphi\varphi' $\\
$\langle (\mu t+\varepsilon)\p_t+(\mu x-t)\p_x+u\p_u\rangle$ & $\varphi(\omega)(\mu t+\varepsilon)^{1/\mu} $
 & $\dfrac{\mu^2x+\varepsilon}{\mu^2(\mu t+\varepsilon)}+\dfrac1{\mu^2}\ln|\mu t+\varepsilon|$
 & $\dfrac1\mu\varphi-\mu\omega\varphi'=(\varphi^{\mu}\varphi)'+\ln\varphi\varphi'$\\
\hline
\end{tabular}
\end{center}}

If $u=f(t,x)$ is an exact solution of equation $\boldsymbol{u_t=(u^{\mu}u_x)_x+\ln u u_x}$, then
\[
u=e^{\varepsilon_1}f\big(e^{\mu\varepsilon_1}t+\varepsilon_2,e^{\mu\varepsilon_1}(x-\varepsilon_1t+\varepsilon_3)\big)
\]
is also solution of the same equation.

In particular, if $\mu=0$, we obtain \mathversion{bold}$u_t=u_{xx}+\ln u u_x$\mathversion{normal}.
This equation has two known travelling wave solutions (usual and generalized ones):
\[
u(t,x)=\exp(c_1e^{-x+c_2t}+1-c_2),\quad u(t,x)=\exp\left(\dfrac{c_1-x}{t+c_2}+\dfrac{\ln|t+c_2|}{t+c_2}\right).
\]

Generalized travelling wave solution are known for the following equations:
\begin{gather*}\textstyle
\mbox{\mathversion{bold}$u_t=((u^{2\mu}+bu^\mu)u_x)_x+u^\mu u_x$,\mathversion{normal}}\quad
u=\left(x\varphi(t)+c_1\varphi(t)+\dfrac bn\varphi(t)\int\varphi(t)dt\right)^{1/\mu}.
\\
\textstyle
\mbox{\mathversion{bold}$u_t=((ae^{2u}+be^u)u_x)_x+e^uu_x$},\quad
   u =\ln\bigl(x\varphi(t)+c_1\varphi(t)+b\varphi(t)\int\varphi(t)dt\bigr),
\end{gather*}
where $\varphi=\varphi(t)$ is determined by the equation $\varphi'=a\varphi^3+\varphi^2$.

\section[$n$-dimensional radially symmetric nonlinear diffusion equations]
{$\boldsymbol{n}$-dimensional radially symmetric nonlinear\\ diffusion equations}\label{SectionRadSym}

Class~\eqref{eqDKfgh} contains a subclass of physically important
$n$-dimensional radially symmetric nonlinear diffusion equations.
Preserving the common terminology we use for them the notation
\begin{equation}\label{eqRadSymNDEs}
u_t=r^{1-n}(r^{n-1}A(u)u_r)_r.
\end{equation}
Equations of the form~\eqref{eqRadSymNDEs}, especially for power nonlinearity
\begin{equation}\label{eqRadSymNDEsPower}
\boldsymbol{u_t=r^{1-n}(r^{n-1}u^\mu u_r)_r.}
\end{equation}
have a large number of applications,
for both $\mu > 0$ (slow diffusion) and  $\mu < 0$ (fast diffusion).

To the best of our knowledge, first similar solutions of~\eqref{eqRadSymNDEs}
were instantaneous source-type solutions~\cite{Barenblatt1952,Pattle1959}
namely, solutions of form
\[
u=t^{-n/(\mu n+2)}f(\xi), \quad \xi=rt^{-1/(\mu n+2)}
\]
if $\mu\ne -2/N$ and
\[
u=e^{-\lambda nt}f(\xi), \quad \xi=re^{\lambda t},
\]
where $\lambda$ is an arbitrary constant, in case $\mu= -2/N$.
Substituting these values into~\eqref{eqRadSymNDEsPower} yields
(assuming $t>0$ in the first three cases):
\[
\begin{array}{ll}
\mu>0:\quad & u=\left\{
\begin{array}{ll}
t^{-n/(\mu n+2)}\Big[\frac\mu{2(\mu n+2)}(a^2-r^2t^{-2/(\mu n+2)})\Big]^{1/\mu},\ & r<at^{1/{\mu n+2}}\\
0,& r\ge at^{1/{\mu n+2}}
\end{array}
\right.,
\\[1ex]
\mu=0:\quad & u=At^{-n/2}e^{-r^2/4t},
\\[1ex]
0>\mu>-2/n:\quad & u=t^{-n/(\mu n+2)}\Big[\frac{-\mu}{2(\mu n+2)}(a^2+r^2t^{-2/(\mu n+2)})\Big]^{1/\mu},
\\[1ex]
\mu=-2/n:\quad & u=e^{-\lambda nt}\Big[\frac\lambda n(a^2+r^2e^{-2\lambda t})\Big]^{-n/2},
\\[1ex]
\mu<-2/n:\quad & u=\left\{\begin{array}{ll}
(-t)^{-n/(\mu n+2)}\Big[\frac\mu{2(\mu n+2)}(a^2+r^2(-t)^{-2/(\mu n+2)})\Big]^{1/\mu},\ & t<0\\
0,& t\ge 0
\end{array}.
\right.
\end{array}
\]

The following instantaneous source-type solutions for equations~\eqref{eqRadSymNDEsPower}
were obtained by King~\cite{King1990}:

Case $\mu=-1$, $n\ne-2$: \quad
$
f=(\mu n+2)\exp\Big(-\frac{\alpha\xi^{-2-n}}{2-n}\Big)/\int\xi\exp\Big(-\frac{\alpha\xi^{-2-n}}{2-n}\Big)d\xi.
$\\

Case $\mu=-1$, $n=1$: \quad $f=1/(\beta e^{-\alpha\xi}-\alpha^{-2}(\mu n+2)^{-1}(1+\alpha\xi))$.\\

Case $\mu=-1$, $n=2$: \quad $f=1/(\beta\xi^\alpha+\xi^2(\mu n+2)^{-1}(2-\alpha)^{-1})$, if $\alpha\ne2$ and\\
\phantom{ Case $\mu=-1$, $n=2$: \quad ~~}{} $f=1/(\beta\xi^2+\xi^2(\mu n+2)^{-1}\ln\xi$, if $\alpha=2$.\\

Case $\mu=-1$, $n=3$:  \quad
$f=2/(2\beta e^{-\alpha/\xi}+(\xi(\alpha+\xi)+\alpha^{-2}e^{-\alpha/\xi}E_1(-\alpha/\xi))(\mu n+2)^{-1})$,
where $E_1(z)=\int_z^\infty e^{-t}t^{-1}dt$ is the exponential integral.\\

Case $\mu=-2/n$: \quad $f=\xi^{-n}g$, where $\int(ng-\alpha g^{2/n}-\lambda g^{1+2/n})dg=\ln\xi$.\\

Case $\mu=-1/2$:
\begin{gather*}
f=\alpha(\mu n+2)\xi^{-n}\left(
\beta J_{\nu-1}\Big(\frac{\alpha^{1/2}\xi^{(4-n)/2}}{(\mu n+2)^{1/2}(4-n)}\Big)
+(1-\beta)Y_{\nu-1}\Big(\frac{\alpha^{1/2}\xi^{(4-n)/2}}{(\mu n+2)^{1/2}(4-n)}\Big)\right)^2\\
\times \left(
\beta J_{\nu}\Big(\frac{\alpha^{1/2}\xi^{(4-n)/2}}{(\mu n+2)^{1/2}(4-n)}\Big)
+(1-\beta)Y_{\nu}\Big(\frac{\alpha^{1/2}\xi^{(4-n)/2}}{(\mu n+2)^{1/2}(4-n)}\Big)
\right)^{-2},
\end{gather*}
where $\nu=2/(4-n)$, $J_\nu(z)$ and $Y_\nu(z)$ are Bessel functions of the first and the second type correspondingly.

We now consider solutions~\cite{King1990} that generalize
the one-dimensional dipole solutions given for $\mu>0$ in~\cite{Barenblatt&Zel'dovich1957}.
The similarity variables are chosen to
fix $\int_o^\infty rudr$ in time (if the integral exists). In one dimension this corresponds to the centre of mass.
The appropriate similarity solution to~\eqref{eqRadSymNDEsPower} then takes the form
\begin{gather*}
u=t^{-1/(\mu+1)}f(\xi), \quad \xi=rt^{-1/(2(\mu +1))}, \quad n\ne-1,
\\
u=e^{-2\lambda t}f(\xi), \quad \xi=re^{-\lambda t}, \quad n=-1.
\end{gather*}
For different values $n$ and $\mu$ the following solutions are known~\cite{Barenblatt&Zel'dovich1957,King1990}:

Case $\mu\ne0,-2/n$:\quad
$f=\xi^{(2-n)/(\mu+1)}\left(\frac{n}{2(\mu n+2)}(\beta-\xi^{(\mu n+2)/(\mu+1)})\right)^{1/n}$.\\

Case $\mu=0$:\quad $f=\beta\xi^{2-n}e^{-\xi^2/4}$.\\

Case $\mu=-2/n$:\quad $f=\left(\frac{\xi^2}{n-2}\ln\frac\xi\beta\right)^{-n/2}$.\\

Case $\mu=-1/2$: (here $\nu=2-n/2$)
\[
f=\frac\alpha{\xi^2}
\left[\beta J_{\nu-1}\Big(\frac{\alpha^{1/2}\xi}2\Big)+(1-\beta)Y_{\nu-1}\Big(\frac{\alpha^{1/2}\xi}2\Big)\right]^2
\left[\beta J_{\nu}\Big(\frac{\alpha^{1/2}\xi}2\Big)+(1-\beta)Y_{\nu}\Big(\frac{\alpha^{1/2}\xi}2\Big)\right]^{-2}.
\]

Case $\mu=-1/2$, $n=3$ (cr \cite{Yang&Chen&Zheng&Pan1990}):\quad
$f=\frac\alpha{\xi^2}\tan^2\Big(\frac{\alpha^{1/2}}2(\xi-c)\Big)$ or
$f=\frac\alpha{\xi^2}\tanh^2\Big(\frac{\alpha^{1/2}}2(\xi-c)\Big)$.\\

Case $\mu=-1/2$, $n=1$:\quad
$f=\frac{\alpha^2}{(2-\alpha^{1/2}\xi\cot[\alpha^{1/2}(\xi-c)/2])^2}$ or
$f=\frac{\alpha^2}{(2-\alpha^{1/2}\xi\coth[\alpha^{1/2}(\xi-c)/2])^2}$.\\

Case $\mu=n/2-2$:\quad $f=\xi^{(2-n)/(\mu+1)}g$, where
\[
\int\frac{g^{n/2-2}}{g+(n-2)\alpha}dg=-\frac1{(n-2)^2}\xi^{n-2}.
\]

\section{Variable coefficient diffusion--convection equations}\label{SectionOnSolutionsOfNVCDCE}

To obtain invariant solutions of the variable coefficient diffusion--convection equations of form~\eqref{eqDKfgh}
two approaches were used. The first one is the direct finding of solutions invariant with respect to a subalgebra of
the Lie invariance algebra, and the second one is reconstructing of new solutions from the known ones
using equivalence transformations.

As an example of implementation of the first approach we adduce some of the invariant solutions
of equation~\cite{Hill1989,Polyanin&Zaitsev2004}.
\mathversion{bold}\begin{equation}\label{EqNDExpum}
|x|^pu_t=(|u|^\mu u_x)_x.
\end{equation}\mathversion{normal}
Namely,
\begin{gather*}
u=|c_1x+c_0|^{1/(\mu+1)},\quad
u=\left(\frac{-c_1\mu}{(p+2)(2+\mu+p+\mu p)}\right)^{1/\mu}(c_1t+c_0)^{-1/\mu}x^{(2+p)/\mu}\\
u=t^{(p+1)\beta}\left(\frac{\mu\beta}{2+p}x^{2+p}t^{(p+2)\beta}+c_0\right)^{1/\mu},
\quad\mbox{where}\quad \beta=-\frac1{p\mu+p+\mu+2}.
\end{gather*}

For equation~\eqref{EqNDExpum} with $p=-(\mu+2)/(\mu+1)$ solution obtained with multiplicative separation of variables
is known:
\[
u=e^{-c_1t}\left(c_1(\mu+1)^2x^{\mu/(\mu+1)}e^{c_1\mu t}+c_2\right)^{1/\mu}
\]

See also the next two sections for invariant solutions of
essentially variable coefficient equations.

Another possible way of finding exact solutions is based on application of equivalence transformations.
The complete extended equivalence group $\hat G^{\Equiv}$ of class~\eqref{eqDKfgh} is
formed by the transformations~\cite{Ivanova&Popovych&Sophocleous2004,Ivanova&Popovych&Sophocleous2007Part1}
\begin{gather}\nonumber
\tilde t=\delta_1 t+\delta_2,\quad
\tilde x=X(x), \quad
\tilde u=\delta_3 u+\delta_4, \\
\tilde f=\dfrac{\varepsilon_1\delta_1\varphi}{X_x}f, \quad
\tilde g=\varepsilon_1\varepsilon_2^{-1}X_x\varphi\,g, \quad
\tilde h=\varepsilon_1\varepsilon_3^{-1}\varphi\,h, \quad
\tilde A=\varepsilon_2 A, \quad
\tilde B=\varepsilon_3 (B+\varepsilon_4 A),\label{EquivGroup}
\end{gather}
where $\delta_j$ $(j=\overline{1,4})$ and $\varepsilon_i$ $(i=\overline{1,4})$ are arbitrary constants,
$\delta_1\delta_3\varepsilon_1\varepsilon_2\varepsilon_3\not=0$,
$X$ is an arbitrary smooth function of~$x$, $X_x\not=0$,
$\varphi=e^{-\varepsilon_4\int \frac{h(x)}{g(x)}dx}$.

It appears also, that class~\eqref{eqDKfgh} contains equations being mutually equivalent with respect to
point transformations which do not belong to this group.
In particular, it is proved
in~\cite{Ivanova&Popovych&Sophocleous2007Part1,Ivanova&Sophocleous2006,Popovych&Ivanova2004NVCDCEs} that
if an equation of form~\eqref{eqDKfgh} is invariant with respect to a Lie algebra
of dimension not less than 4 then it can be reduced by point transformations to a constant coefficient
diffusion--convection equation~\eqref{eqCCDCE}.
All such equations and corresponding transformations
were found in~\cite{Ivanova&Popovych&Sophocleous2007Part1,Ivanova&Sophocleous2006,Popovych&Ivanova2004NVCDCEs}.
Some of them were known previously~\cite{Munier&Burgan&Gutierres&Fijalkow&Feix1981}.
For the convenience of the readers we adduce the results of group classification up to the
extended equivalence group~\eqref{EquivGroup} in Appendix~\ref{AppendixGroupClassif}.

Up to equivalence transformations~\eqref{EquivGroup} the list of equations of form~\eqref{eqDKfgh}
reducible to the constant coefficient
form together with corresponding transformations
is exhausted by the following ones~\cite{Ivanova&Popovych&Sophocleous2007Part1}:
\bigskip

\setcounter{casetran}{0}
\noindent
\refstepcounter{casetran}\thecasetran. $u_t=(e^uu_x)_x+\varepsilon xu_x$ $\to$ \eqref{solv1}:\quad
$\tilde t=e^{2\varepsilon t}/(2\varepsilon)$, $\tilde x=xe^{\varepsilon t}$, $\tilde u=u$;
\\[1ex]
\refstepcounter{casetran}\thecasetran.
$x^{-3}u_t=(e^uu_x)_x$ $\to$ \eqref{solv1}:\quad $\tilde t= t\sign x$, $\tilde x=1/x,$ $\tilde u=u-\ln |x|$;
\\[1ex]
\refstepcounter{casetran}\thecasetran. $x^{-3}u_t=(e^uu_x)_x+x^{-2}u_x$ $\to$ \eqref{solv1}:\quad
$\tilde t=(e^{2t}t\sign x)/2$,
$\tilde x=e^{-t}/x,$ $\tilde u=u-t-\ln |x|$;
\\[1ex]
\refstepcounter{casetran}\thecasetran. $u_t=(|u|^\mu u_x)_x+\varepsilon xu_x$ $\to$ \eqref{solv5}:\quad
$\tilde t=e^{2\varepsilon t}/(2\varepsilon)$, $\tilde x=xe^{\varepsilon t}$, $\tilde u=u$;
\\[1ex]
\refstepcounter{casetran}\thecasetran.
$|x|^{-\frac{3\mu+4}{\mu+1}}u_t=(|u|^\mu u_x)_x$ $\to$ \eqref{solv5}:\quad
$\tilde t=t$, $\tilde x=-1/x$, $\tilde u=|x|^{-\frac1{1+\mu}}u$;
\\[1ex]
\refstepcounter{casetran}\thecasetran.
$|x|^{-\frac{3\mu+4}{\mu+1}}u_t=(|u|^\mu u_x)_x+\varepsilon x|x|^{-\frac{3\mu+4}{\mu+1}}u_x\big|_{\mu\ne2}$ $\to$
\eqref{solv5}:\quad
$\tilde t=\dfrac{\mu+1}{\varepsilon(\mu+2)}(1-e^{-\varepsilon\frac{\mu+2}{\mu+1}t})$, $\tilde x=xe^{\varepsilon t}$,\\
\phantom{\thecasetran.}$\tilde u=u$;
\\[1ex]
\refstepcounter{casetran}\thecasetran. $x^{-2}u_t=(u^{-2}u_x)_x+\varepsilon x^{-1}u_x$ $\to$ \eqref{eqFuSE}:\quad
$\tilde t=t$, $\tilde x=xe^{\varepsilon t}$, $\tilde u=u$;
\\[1ex]
\refstepcounter{casetran}\thecasetran.
$e^xu_t=(u^{-1}u_x)_x$ $\to$ \eqref{eqFDE}:\quad $\tilde t=t$, $\tilde x=x$, $\tilde u=e^{x}u$;
\\[1ex]
\refstepcounter{casetran}\thecasetran. $e^xu_t=(u^{-1}u_x)_x+\varepsilon e^xu_x$ $\to$ \eqref{eqFDE}:\quad
$\tilde t=e^{\varepsilon t}/\varepsilon$, $\tilde x=x+\varepsilon t$, $\tilde u=e^{x+\varepsilon t}u$.
\\[1ex]

Combining these transformations with symmetry and equivalence transformations one can easily obtain solutions of
such ``non-essentially variable coefficient" equations.
For instance, starting from solutions of equation~\eqref{solv1}, we obtain
corresponding solutions for the more complicated and interesting equation
\mathversion{bold}\[
\dfrac{e^x}{(\gamma e^x+1)^3}\,u_t=(e^uu_x)_x+e^uu_x
\]\mathversion{normal}
having the density $f$ localized in space~\cite{Popovych&Ivanova2004NVCDCEs}:
\[
u=\ln\left|c_1+c_0(e^{-x}+\gamma)\right|, \quad
u=\ln\left(-\frac1{2t(e^{-x}+\gamma)}-\frac{c_1}t+c_0\frac{e^{-x}+\gamma}t\right).
\]
Similarly one can find exact solutions of equation
\mathversion{bold}\[
e^xu_t=(u^{-1}u_x)_x+\varepsilon e^xu_x
\]\mathversion{normal}
that is reducible to the fast diffusion equation~\cite{Ivanova&Popovych&Sophocleous2007Part2}:
\begin{gather*}
u=\dfrac{e^{-(x+\varepsilon t)}}{1+c e^{x+\varepsilon t+e^{\varepsilon t}/\varepsilon}},\qquad
u=\dfrac{\varepsilon e^{-(x+\varepsilon t)}}{\varepsilon x+\varepsilon^2t-e^{\varepsilon t}
\pm e^{\varepsilon t-\varepsilon(x+\varepsilon t)e^{-\varepsilon t}}},
\\[1ex]
u=\dfrac{2e^{-x}}{\varepsilon(x+\varepsilon t)^2\pm e^{2\varepsilon t}},\qquad
u=\dfrac{2e^{-x}}{\varepsilon\cos^2(x+\varepsilon t)},\qquad u=-\dfrac{2e^{-x}}{\varepsilon\cosh^2(x+\varepsilon t)},
\\[1ex]
u=\dfrac{2e^{-x}}{\varepsilon\sin^2(x+\varepsilon t)},\qquad
u=\dfrac{2e^{-(x+\varepsilon t)}\sin(2e^{\varepsilon t}/\varepsilon)}{\cos(2e^{\varepsilon t}/\varepsilon)
-\cos2(x+\varepsilon t)},
\\[1ex]
u=\dfrac{2e^{-(x+\varepsilon t)}\sinh(2e^{\varepsilon t}/\varepsilon)}{\cosh2(x+\varepsilon t)
-\cosh(2e^{\varepsilon t}/\varepsilon)},\quad
u=-\dfrac{2e^{-(x+\varepsilon t)}\sinh(2e^{\varepsilon t}/\varepsilon)}{\cosh2(x+\varepsilon t)
+\cosh(2e^{\varepsilon t}/\varepsilon)},
\\[1ex]
u=\dfrac{2e^{-(x+\varepsilon t)}\cosh(2e^{\varepsilon t}/\varepsilon)}{\sinh2(x+\varepsilon t)
-\sinh(2e^{\varepsilon t}/\varepsilon)},\quad
u=\dfrac{2e^{-(x+\varepsilon t)}\sin(2e^{\varepsilon t}/\varepsilon)}{\cosh2(x+\varepsilon t)
-\cos(2e^{\varepsilon t}/\varepsilon)},
\\[1ex]
u=\dfrac{2e^{-(x+\varepsilon t)}\sinh(2e^{\varepsilon t}/\varepsilon)}{\cosh(2e^{\varepsilon t}/\varepsilon)
-\cos2(x+\varepsilon t)}.
\end{gather*}

Using the same approach we constructed exact solutions for the following
equations~\cite{Ivanova&Popovych&Sophocleous2007Part2,Ivanova&Sophocleous2006,Popovych&Ivanova2004NVCDCEs}:
\[
\boldsymbol{e^{-2x+\gamma e^{-x}}u_t=(u^{-1}u_x)x+u^{-1}u_x}:
\]
\begin{gather*}
u=c_0e^{(c_1-\gamma)e^{-x}},\qquad
u=\dfrac{2c_1^2te^{-\gamma e^{-x}}}{\cos^2c_1(e^{-x}+c_0)},\qquad
u=\dfrac{2tc_0c_1^2e^{(c_1-\gamma)e^{-x}}}{(1-c_0e^{c_1e^{-x}})^2},
\\
u=\dfrac{c_1e^{-\gamma e^{-x}}}{-\varepsilon+c_0e^{c_1(e^{-x}-\varepsilon t)}},\qquad
u=\frac{\varepsilon e^{-\gamma e^{-x}}}
{e^{-x}-\varepsilon t+c_0}, \qquad
u=\frac{2te^{-\gamma e^{-x}}}{(e^{-x}+c_1)^2+c_0t^2}.
\end{gather*}
Equation
\[
\boldsymbol{\dfrac{e^{-2x}}{(e^{-x}+\gamma)^{\frac{4+3\mu}{1+\mu}}}\,u_t=\left(u^{\mu}u_x\right)_x+u^{\mu}u_x}
\]
has exact solutions of the form
\begin{gather*}
u=|c_0(e^{-x}+\gamma)-c_1|^{\frac1{\mu+1}},
\quad
u=\left(c_0+\frac{\varepsilon\mu}{e^{-x}+\gamma}+\varepsilon^2\mu t\right)^{\frac1\mu}
|e^{-x}+\gamma|^{-\frac1{\mu+1}},
\\
u=\left(-\frac\mu{\mu+2}\,\frac1{2t}\left(c_0-\frac1{e^{-x}+\gamma}\right)^2
+c_1|t|^{-\frac{\mu}{\mu+2}}\right)^{\frac1\mu}
|e^{-x}+\gamma|^{-\frac1{\mu+1}},
\\
u=\left(-\frac\mu{\mu+2}\,\frac1{2t}\left(c_0-\frac1{e^{-x}+\gamma}\right)^2
+c_1\left(c_0-\frac1{e^{-x}+\gamma}\right)^{\frac\mu{\mu+1}}
|t|^{-\frac{\mu(2\mu+3)}{2(\mu+1)^2}}
\right)^{\frac1\mu}|e^{-x}+\gamma|^{-\frac1{\mu+1}},
\\
u=(6t+c_1'+c_2e^{-x})^2(e^{-x}+\gamma)^6.
\end{gather*}

\[
\boldsymbol{e^xu_t=(u^{-1}u_x)_x+\mu e^xu_x}
\]
admits the following invariant exact solutions:
\begin{gather*}
u=\frac{c_0}{\mu}e^{(1-c_1)(x+\mu t)},\quad u=\dfrac{2c_1^2}{\mu}\dfrac{e^{-x}}{\cos^2[c_1(x+\mu t+c_0)]},\quad
u=\dfrac{2c_0c_1^2}{\mu}\dfrac{e^{(c_1-1)x+c_1\mu t}}{[1-c_0 e^{c_1(x+\mu t)}]^2},\\
u=\dfrac{c_1}{\mu}\dfrac{e^{-(x+\mu t)}}{c_2+c_0e^{c_1(x+\mu t+ c_2e^{\mu t})}},\quad
u=\dfrac{2}{\mu}\dfrac{e^{-x}}{(x+\mu t+c_1)^2+c_0e^{2\mu t}},\\
u=\dfrac 1{\mu}\dfrac{e^{-(x+\mu t)}}{x+\mu t-e^{\mu t}+c_0}.
\end{gather*}

\[
\boldsymbol{f(x)u_t=(u^{-1}u_x)_x-\left (\frac{f_x}{2f}-
\frac{\sqrt f}{\int\sqrt fd x}\right )u^{-1}u_x},
\]
where $f=f(x)$ is an arbitrary positive function has the following invariant solutions:
\begin{gather*}
u=c_0(\int\sqrt{f}d x)^{c_1-2},\quad u=\dfrac{2c_1^2t(\int\sqrt{f}d x)^{-2}}{\cos^2[c_1(\ln\int\sqrt{f}d x+c_0)]},\quad
u=\dfrac{2c_0c_1^2t(\int\sqrt{f}d x)^{c_1-2}}{[1-c_0(\int\sqrt{f}d x)^{c_1}]^2},\\
u=\dfrac{c_1(\int\sqrt{f}d x)^{-2}}{c_2+c_0(\int\sqrt{f}d x)^{c_1}e^{c_1c_2t}},\quad
u=\dfrac{(\int\sqrt{f}d x)^{-2}}{\ln (\int\sqrt{f}d x)-t+c_0},\\
u=\dfrac{2t(\int\sqrt{f}d x)^{-2}}{[\ln (\int\sqrt{f}d x)+c_1]^2+c_0t^2}.
\end{gather*}

\[
\boldsymbol{u_t=(u^{-1}u_x)_x+\mu xu_x}
\]
admits the following solutions:
\begin{gather*}
u=c_0e^{c_1xe^{\mu t}},\quad u=\dfrac{c_1^2e^{2\mu t}}{\mu\cos^2[c_1(xe^{\mu x}+c_0)]},\quad
u=\dfrac{c_0c_1^2e^{(2\mu t+c_1xe^{\mu t})}}{\mu (1-c_0e^{c_1xe^{\mu t}})^2},\\
u=\dfrac{c_1}{c_2+c_0e^{c_1(xe^{\mu t}+\dfrac{c_2}{2\mu}e^{2\mu t})}},\quad
u=\dfrac{2\mu}{2\mu xe^{\mu t}-e^{2\mu t}+2\mu c_0},\\
u=\dfrac{4\mu e^{2\mu t}}{4\mu^2(xe^{\mu t}+c_1)^2+c_0e^{4\mu t}}.
\end{gather*}

See~\cite{Ivanova&Popovych&Sophocleous2007Part1,Ivanova&Popovych&Sophocleous2007Part2,Ivanova&Sophocleous2006,
Popovych&Ivanova2004NVCDCEs}
for more detail and more examples.
%Definitely the same trick can be used for obtaining non-Lie solutions starting
%from ones of constant coefficient equations.
Since the solutions of these equations can be reconstructed from
ones presented in other sections, we tern back to the more interesting cases, in particular, to the equations
which are ``essentially variable coefficient''.

\section{Examples of Lie reduction of variable coefficient equation}\label{SectionOnExactSolOfGenBurEq}

The results of this section have been obtained in~\cite{Ivanova&Popovych&Sophocleous2007Part2,Ivanova&Sophocleous2006}.

In this section we consider Lie reductions of some variable coefficient equations
of form~\eqref{eqDKfgh}. We start from the equation
\begin{equation}\label{e1}
\boldsymbol{e^{px}u_t=\left [e^uu_x\right ]_x+e^{px}u_x}
\end{equation}
which admits three-dimensional Lie algebra
$Q_1=-p^{-1}\p_t$, $Q_2=e^{-pt}(\p_t-\p_x)$, $Q_3=\p_x+p\p_u$.
The only non-zero commutators of these operators are
$[Q_1,Q_2]=Q_2.$ Therefore $A^{\max}$ is a realization
of the algebra~$2A_{2.1}$~\cite{Mubarakzyanov1963}.
All the possible inequivalent (with respect to inner automorphisms)
one-dimensional subalgebras of~$A_{2.1}\oplus A_1$~\cite{Patera&Winternitz1977} are exhausted by the ones
listed in Table~\ref{TableSelfSimSolForNDCEe1} along with
the corresponding ans\"atze and the reduced odes.

%\bigskip
{\begin{center}\refstepcounter{tabul}\label{TableSelfSimSolForNDCEe1}
Table~\thetabul. Reduced odes for (\ref{e1}). $\lambda\ne0,$ $\varepsilon=\pm1$
\\[1.5ex] \footnotesize
\begin{tabular}{|l|l|c|c|l|}
\hline
N&Subalgebra& Ansatz $u=$& $\omega$ &\hfill {Reduced ODE\hfill} \\
\hline
1&$Q_1$&$\varphi (\omega )$&$x$&$\left (e^{\varphi}\varphi'\right )'
+e^{px}\varphi'=0$  \\
2&$Q_2$&$\varphi (\omega )$&$t+x$&$\left (e^{\varphi}\varphi'\right )'=0$ \\
3&$Q_3$&$\varphi (\omega )+px$&$t$&$\varphi'=p^2e^{\varphi}+p$ \\
4&$Q_1+\lambda Q_3$&$\varphi (\omega )+\lambda pt$&$x-\lambda t$&
$e^{\omega}\left [\lambda p-(\lambda +1)\varphi'\right ]=
\left (e^{\varphi}\varphi'\right )'$ \\
5&$Q_2+\varepsilon Q_3$&$\varphi (\omega )+\varepsilon e^{pt}$&$x+t-\frac{\varepsilon}pe^{pt}$&
$e^{p\omega}\varepsilon (p-\varphi')=\left (e^{\varphi}\varphi' \right )'$  \\
\hline
\end{tabular}
\end{center}

\medskip

As a second example we consider the equation
\begin{equation}\label{e2}
\boldsymbol{x^{p}u_t=\left [u^mu_x\right ]_x+x^{p+1}u_x.}
\end{equation}
The invariance algebra of~(\ref{e2}) is generated by the operators
$Q_1=-(p+2)^{-1}\p_t,$ $Q_2=e^{-(p+2)t}(\p_t-x\p_x),$ $Q_3=mx\p_x+(p+2)u\p_u $
and is a realization of the algebra~$A_{2.1}\oplus A_1$ too. The reduced equations for~(\ref{e2})
are listed in table~\ref{TableSelfSimSolForNDCEe2}.

{\begin{center}\refstepcounter{tabul}\label{TableSelfSimSolForNDCEe2}
Table~\thetabul. Reduced odes for (\ref{e2}). $\lambda\ne0,$ $\varepsilon=\pm1$
\\[1.5ex] \footnotesize
\begin{tabular}{|l|l|c|c|l|}
\hline
N&Subalgebra& Ansatz $u=$& $\omega$ &\hfill {Reduced ODE\hfill} \\
\hline
1&$Q_1$&$\varphi (\omega )$&$x$&$\left ({\varphi}^m\varphi'\right )'+x^{p+1}\varphi'=0$  \\
2&$Q_2$&$\varphi (\omega )$&$xe^t$&$\left (\varphi^m\varphi'\right )'=0$ \\
3&$Q_3$&$x^{\frac{p+2}m}\varphi (\omega )$&$t$&$\varphi'=
\frac{(p+2)(p+2+m)}{m^2}\varphi^{m+1}+\frac{p+2}m\varphi$ \\
4&$Q_1+\lambda Q_3$&$e^{\lambda (p+2)t}\varphi (\omega )$&$xe^{-\lambda mt}$&
$\left (\varphi^m\varphi'\right )'+(1+\lambda m)\omega^{p+1}\varphi'$\\&&&&$=
\lambda (p+2)\omega^p\varphi$ \\
5&$Q_2+\varepsilon Q_3$&$e^{\varepsilon e^{(p+2)t}}\varphi (\omega )$&
$xe^{t-\frac{m\varepsilon}{p+2}e^{(p+2)t}}$&
$\left (\varphi^m\varphi'\right )'=\varepsilon (p+2)\omega^p\varphi$\\&&&&$ -m\varepsilon\omega^{p+1}
\varphi'$ \\
\hline
\end{tabular}
\end{center}
%\large\smallskip

At last, let us analyze in more detail equation
\begin{equation}\label{Eqa1bufeax2ghheax2}
\boldsymbol{e^{px^2}u_t=(e^{px^2}u_x)_x+e^{px^2}uu_x,}
\end{equation}
which is invariant with respect to three-dimensional Lie symmetry algebra
\[
\langle\p_t,\,e^{-2pt}\p_x,\,\p_x-2p\p_u\rangle.
\]
In contrast to the case of equations with four-dimensional Lie symmetry algebra
we cannot reduce equation~\eqref{Eqa1bufeax2ghheax2} to a constant coefficient equation
of form~\eqref{eqDKfgh}.
However, it is an interesting feature of this equation that using a point transformation
$v=u+2px$ it can be mapped to a constant coefficient reaction--convection--diffusion equation
\begin{equation}\label{EqCCa1bufeax2ghheax2}
v_t=v_{xx}+vv_x-2pv
\end{equation}
that does not belong to class~\eqref{eqDKfgh}.
For simplification of the technical calculations
we will investigate the constant coefficient equation~\eqref{EqCCa1bufeax2ghheax2} instead of~\eqref{Eqa1bufeax2ghheax2}.
The Lie symmetry algebra of equation~\eqref{EqCCa1bufeax2ghheax2}
\[
\langle X_1=\p_t,\,X_2=e^{-2pt}(\p_x+2p\p_v),\,X_3=\p_x\rangle
\]
is a realization of $A_{2.1}\oplus A_1$~\cite{Mubarakzyanov1963}.
These operators generate the following group of point transformations:
\[
\tilde t=t+\varepsilon_1,\quad \tilde x=x+\varepsilon_2e^{-2pt}+\varepsilon_3,\quad \tilde v=v+2\varepsilon_2pe^{-2pt}.
\]
A list of proper inequivalent subalgebras of the given algebra is exhausted by the following ones
\[
\langle X_1+\alpha X_3\rangle,\quad \langle X_3+\varepsilon X_2\rangle,\quad \langle X_2\rangle,\quad
\langle X_1,\,X_3\rangle,\quad \langle X_2,\,X_3\rangle,\quad \langle X_1+\beta X_3,\,X_2\rangle,
\]
where $\alpha$ and $\beta$ are arbitrary constants, $\varepsilon=0,\pm1$~\cite{Patera&Winternitz1977}.

The first three (one-dimensional) subalgebras lead to Lie reductions to ordinary differential equations,
the fourth and sixth (two-dimensional) ones yield reductions to algebraic equations.
Lie reductions with respect to these subalgebras are summarized in Table~\ref{TableReductionOfGenBurgEq}.
One can easily check that it is impossible to construct
a Lie ansatz corresponding to the subalgebra $\langle X_2,\,X_3\rangle$.

{\begin{center}\refstepcounter{tabul}\label{TableReductionOfGenBurgEq}
Table~\thetabul. Lie reductions of equation~\eqref{EqCCa1bufeax2ghheax2}.
\\[1.5ex] \footnotesize
\begin{tabular}{|l|l|c|c|l|}
\hline \vspacebefore
N&Subalgebra& Ansatz $v=$& $\omega$ &\hfill {Reduced equation\hfill} \\
\hline \vspacebefore
1&$\langle X_1+\alpha X_3\rangle$ & $\varphi(\omega)$ & $x-\alpha t$
 & $\varphi''+(\varphi+\alpha)\varphi'-2p\varphi=0$\\[0.5ex]
 \vspacebefore
2&$\langle X_3+\varepsilon X_2\rangle$ &  $\varphi(\omega)+\dfrac{2p\varepsilon x}{e^{2pt}+\varepsilon}$ & $t$
& $\varphi'=\dfrac{2pe^{2p\omega}}{e^{2p\omega}+\varepsilon}\varphi$\\[0.7ex]
 \vspacebefore
3&$\langle X_2\rangle$ & $\varphi(\omega)+2px$ & $t$
& $\varphi'=0$\\[0.5ex]
 \vspacebefore
4&$\langle X_1,\,X_3\rangle$ & $C$ & --- & $C=0$\\[0.5ex]
 \vspacebefore
5&$\langle X_1+\beta X_3,\,X_2\rangle$ & $2px-2p\beta t+C$ & --- & $-2p\beta=0$\\[0.5ex]
\hline
\end{tabular}
\end{center}

Integration of equations 2--5 from Table~\ref{TableReductionOfGenBurgEq}
give the following invariant solutions of equation~\eqref{EqCCa1bufeax2ghheax2}:
\begin{gather*}
v=0,\quad v=2px+C,\quad v=\dfrac{2p\varepsilon x+Ce^{2pt}}{e^{2pt}+\varepsilon}.
\end{gather*}
The corresponding exact invariant solutions of equation~\eqref{Eqa1bufeax2ghheax2} have the form
\begin{gather}\label{solutions.for.Eqa1bufeax2ghheax2}
u=-2px,\quad u=C,\quad u=\dfrac{2p\varepsilon x+Ce^{2pt}}{e^{2pt}+\varepsilon}-2px,
\end{gather}
where $C$ is an arbitrary constant.

Ansatzes~4.2 and~4.3 give a hint for a possible form
\[
v=\varphi(t)x+\psi(t)
\]
of nonlinear separation of variables for construction of exact solutions of equation~\eqref{EqCCa1bufeax2ghheax2}.
Substitution of the ansatz to equation~\eqref{EqCCa1bufeax2ghheax2} leads to antireduction:
\[
\varphi'=\varphi^2-2p\varphi,\quad \psi'=\varphi\psi-2p\psi.
\]
Solving the above system of ODEs for $\varphi$ and $\psi$ we obtain exactly the solutions
of equations~4.2 and~4.3.

\begin{note}
Using the point transformation $\tilde t=e^{-2pt}$, $\tilde x=x$, $\tilde v=e^{2pt}v$
equation~\eqref{EqCCa1bufeax2ghheax2} can be mapped to
a variable coefficient Burgers equation
$\tilde v_{\tilde t}=-2p{\tilde t}^{-1}\tilde v_{\tilde x\tilde x}-2p\tilde v\tilde v_{\tilde x}$
studied in~\cite{Kingston&Sophocleous1991}.
\end{note}

\begin{note}
The well-known Cole--Hopf transformation $v=2w_x/w$ reduces equation~\eqref{EqCCa1bufeax2ghheax2}
to the famous constant coefficient reaction--diffusion equation with weak nonlinearity
\[%begin{equation}\label{EqReactionDifWeakNonlin}
w_t=w_{xx}-2pw\ln|w|.
\]%end{equation}
After application of the Cole--Hopf transformation to the list of known exact solutions
(see, e.g.,~\cite{Polyanin&Zaitsev2004}) of the equation
with weak nonlinearity we obtain exactly solutions~\eqref{solutions.for.Eqa1bufeax2ghheax2}
of equation~\eqref{EqCCa1bufeax2ghheax2}.
\end{note}

\section{Exact solutions of {\mathversion{bold}$sl(2,\mathbb{R})$}-invariant equation}\label{SectionOnSolutionOfSL2REq}

Analyzing the results of group classification of diffusion--convection equations, we can observe a number
of $\hat G^{\Equiv}$-inequivalent equations~\eqref{eqDKfgh}
which are invariant with respect to different realizations of the algebra~$sl(2,\mathbb{R})$.
The set of such equations is practically exhausted by
the well-known (``constant coefficient'') Burgers and $u^{-4/3}$-diffusion equations
and by the equations which are equivalent to them with respect to additional transformations.
This set is supplemented by the unique essentially variable coefficient
equation~\cite{Ivanova&Popovych&Sophocleous2007Part1,Ivanova&Sophocleous2006}
\mathversion{bold}\begin{equation}\label{A-65B1fx2hx2Copy}
x^2u_t=(u^{-6/5}u_x)_x+x^2u_x,
\end{equation}\mathversion{normal}%
$sl(2,\mathbb{R})$-invariance of~\eqref{A-65B1fx2hx2Copy} is directly connected
with the fact that $h$ is not constant.
The corresponding realization
\[
P_t=\p_t,\quad D=2t\p_t+2x\p_x-5u\p_u,\quad \Pi=t^2\p_t+(2tx+x^2)\p_x-5(t+x)u\p_u.
\]
of the algebra~$sl(2,\mathbb{R})$ is quite different from ones of
cases of Burgers and $u^{-4/3}$-diffusion equations and
is the maximal Lie invariance algebra of equation~\eqref{A-65B1fx2hx2Copy}.
It was a reason  to study equation~\eqref{A-65B1fx2hx2Copy} from the symmetry point of view
in detail in~\cite{Ivanova&Popovych&Sophocleous2007Part2}.
These operators generate the following one-parameter groups of point transformations:
\begin{gather*}
P_t\colon \quad \tilde t=t+\varepsilon,\quad \tilde x=x,\quad \tilde u=u;
\\[1ex]
D\colon \quad \tilde t=e^{\varepsilon}t,\quad \tilde x=e^{\varepsilon}x,\quad  \tilde v=e^{3\varepsilon}v;
\\[.5ex]
\Pi\colon \quad  \tilde t=\dfrac{t}{1-\varepsilon t}, \quad
\tilde x=\dfrac{t+x}{1-\varepsilon(t+x)}-\dfrac{t}{1-\varepsilon t},\quad
\tilde u=(1-\varepsilon(t+x))^6u.
\end{gather*}
The complete Lie invariance group~$G^{\max}$ is generated by both the above continuous transformations and
the discrete transformation of changing of sign in the triple~$(t,x,u)$.
The transformations from~$G^{\max}$ can be used for construction of new solutions from known ones.

A list of proper $G^{\max}$-inequivalent subalgebras of~$A^{\max}$ is exhausted by the algebras
$\langle P_t\rangle$, $\langle D\rangle$, $\langle P_t+\Pi\rangle$, $\langle P_t,\, D\rangle$.
Reduction of~\eqref{A-65B1fx2hx2Copy} with respect to these subalgebras
and application of the invariance transformations
lead to the following set of $G^{\max}$-inequivalent
Lie invariant exact solutions (below $\delta\in\{0,1\}$):
\begin{gather*}
u=C(t+x)^{-5},\quad
u=2^{-5/6}\left(\dfrac{x}{t}\right)^{-5/2}(t+x)^{-5/2},\quad
u=\left(\dfrac{3x^4}{4t}\dfrac{(t+x)^2}{Ct-1}+2\dfrac{x^3}{t^3}(t+x)^3\right)^{-5/6},\\[1ex]
u=x^{-5/6}\left(\dfrac{(t+x)^2}{Ct+1}\right)^{-5/6}\left(\dfrac54\dfrac{x^3}{t^3}
+2\dfrac{x^2}{t^2}(C(t+x)+1)\right)^{-5/6}.
\end{gather*}

In~\cite{Ivanova&Popovych&Sophocleous2007Part2} it was proposed to use functional separation of variables
\begin{equation}\label{PolynomXd6}
u=\left(\sum_{i=0}^6 \varphi^i(t)x^i\right)^{-5/6}.
\end{equation}
to obtain solutions of equation~\eqref{A-65B1fx2hx2Copy}.
The set of all solutions of the form~\eqref{PolynomXd6} is closed with respect to transformations
from~$G^{\max}$ and is exhausted, up to translations with respect to~$t$ and scale transformations,
by the above solutions $u=\delta$ and $u=\delta(t+x)^{-5}$ and the solutions given by the generalized ansatz
\begin{equation}\label{A-65B1fx2hx2SuperAnsatz}
u=(2x^3+\varphi^4(t)x^4+\varphi^5(t)x^5+\varphi^6(t)x^6)^{-5/6}
\end{equation}
and the corresponding reduced system
\begin{equation}\label{A-65B1fx2hx2CopyReducedSystem}
\varphi^4_t = 7\varphi^5-\dfrac43(\varphi^4)^2, \quad
\varphi^5_t = 18\varphi^6-\dfrac43\varphi^4\varphi^5, \quad
\varphi^6_t = -\dfrac56(\varphi^5)^2+2\varphi^4\varphi^6.
\end{equation}
System~\eqref{A-65B1fx2hx2CopyReducedSystem} can be reduced to the single third-order ordinary
differential equation on the function~$\varphi^4$:
\[
63\varphi^4_{ttt}+387(\varphi^4_t)^2+126\varphi^4\varphi^4_{tt}+192(\varphi^4)^2\varphi^4_t+16(\varphi^4)^4=0
\]
having particular solutions $\varphi^4=0$, $\varphi^4=C/t$, where $C\in\{0, 3/4, 21/4, 6\}$,
that lead to Lie invariant solutions
of~\eqref{A-65B1fx2hx2Copy}.

\subsection*{Conclusive remarks}
This work is constantly under updating. So, the author will appreciate any suggestions, comments and references
sent to ivanova@imath.kiev.ua.

\subsection*{Acknowledgment}
The author is grateful to G.W.~Bluman, V.M.~Boyko,  V.I.~Lahno, O.I.~Morozov, A.G.~Nikitin, R.O.~Popovych,
A.G.~Sergyeyev, C.~Sophocleous, O.O.~Vaneeva, I.A.~Yehorchenko, O.Yu.~Zhalij
for fruitful discussions and suggested references.
She acknowledges the hospitality and financial support of the Department of Mathematics of the University of British Columbia
and Department of Mathematics and Statistics of the University of Cyprus
where the part of research has been made.
%This research was partially supported by the grant of the President of Ukraine for young scientists
%(project number GP/F11/0061Â).

%\end{document}

\newpage

\begin{appendix}

\section{Group classification of diffusion--convection equations}\label{AppendixGroupClassif}

\begin{center}\footnotesize\renewcommand{\arraystretch}{1.2}\refstepcounter{tabul}
Table~\thetabul. Case of $\forall A(u)$ (gauge $g=1$)\\[1ex]
\begin{tabular}{|l|c|c|c|l|}
\hline
N & $B(u)$ & $f(x)$ & $h(x)$ & \hfil Basis of A$^{\rm max}$ \\
\hline
\refstepcounter{tbn}\label{gcAaBafaha}\thetbn & $\forall$ & $\forall$ & $\forall$ & $\p_t$ \\
%1
\refstepcounter{tbn}\label{gcAaBafexh1}\thetbn a&$\forall$ & $e^{p x}$ & 1 &$\p_t,\, p t\p_t+\p_x$ \\
\thetbn a$'$ & $\forall$ & $|x|^p$ & $x^{-1}$ & $\p_t,\, (p+2)t\p_t+x\p_x$ \\
\thetbn b & 1 & $e^x$ & $e^x+\beta$ & $\p_t,\, e^{-t}(\p_t-\p_x)$ \\
\thetbn c & 1 & $|x|^p$ & $x|x|^p+\beta x^{-1}$ & $\p_t,\, e^{-(p+2)t}(\p_t-x\p_x)$ \\
%2
\refstepcounter{tbn}\label{gcAaB1fx-2hlnxx}\thetbn & 1 & $x^{-2}$ & $x^{-1}\ln|x|$ & $\p_t,\, e^{-t}x\p_x$ \\
%3
\refstepcounter{tbn}\label{gcAaB0f1ha}\thetbn& 0 & 1 & 1 & $\p_t,\, \p_x,\, 2t\p_t+x\p_x$ \\
%4
\hline
\end{tabular}
\end{center}
{\footnotesize

Here $p\in\{0,1\}\!\!\mod G^{\sim}_1$ in case~\ref{gcAaBafexh1};
$p\ne-2$ in case \ref{gcAaBafexh1}c;
%$\beta\in\{0,1\}$ in case \ref{gcAaBafexh1}$'$a;
$\beta\in\{0,\pm 1\}$ in case \ref{gcAaBafexh1}b.
\\
\setcounter{casetran}{0}
Additional equivalence transformations:\\[1ex]
\refstepcounter{casetran}\thecasetran. \ref{gcAaBafexh1}a($p=0$, $B=1$)
$\to$ \ref{gcAaBafexh1}a($p=0$, $B=0$):\quad
$\tilde t=t$, $\tilde x=x+t$, $\tilde u=u$;
\\[1ex]
\thecasetran$'$. \ref{gcAaBafexh1}a$'$($p=-2$, $B=1$) $\to$
\ref{gcAaBafexh1}a$'$($p=-2$, $B=0$):\quad $\tilde t=t$, $\tilde x=xe^t$, $\tilde u=u$;
\\[1ex]
\refstepcounter{casetran}\thecasetran. \ref{gcAaBafexh1}b $\to$ \ref{gcAaBafexh1}a($B=\beta$, $p=1$):\quad
$\tilde t=e^t$, $\tilde x=x+t$, $\tilde u=u$;
\\[1ex]
\refstepcounter{casetran}\thecasetran. \ref{gcAaBafexh1}c($p\ne-2$) $\to$ \ref{gcAaBafexh1}$'$a($p\ne-2$):\quad
$\tilde t=(e^{(p+2)t}-1)/(p+2)$, $\tilde x=xe^t$, $\tilde u=u$.
\par}

%\newpage
\vspace{3ex}

\setcounter{tbn}{0}

\begin{center}\footnotesize\renewcommand{\arraystretch}{1.2}\refstepcounter{tabul}
Table~\thetabul. Case of $A(u)=e^{\mu u}$ \\[1ex]
\begin{tabular}{|l|c|c|c|c|l|}
\hline
N & $B(u)$ & $f(x)$ & $g(x)$ & $h(x)$ & \hfil Basis of A$^{\rm max}$ \\
\hline
\refstepcounter{tbn}\label{AeuB0faha}\thetbn & 0& $\forall$ & 1 & 1 & $\p_t,\, t\p_t-\p_u$ \\
%1
\refstepcounter{tbn}\label{AeuBenufemxhex}\thetbn & $e^{\nu u}$ & $|x|^p$ & 1 & $|x|^q$ &
$\p_t,\, (p\mu-p\nu-2\nu-q\mu+\mu)t\p_t+(\mu-\nu)x\p_x+(q+1)\p_u$ \\
%2
%\refstepcounter{tbn}\label{AeuBenufxmhxl}
\thetbn${}^*$ & $e^{\nu u}$ & $e^{p x}$ & 1 & $\varepsilon e^{x}$ &
$\p_t,\, (p\mu-p\nu-\mu)t\p_t+(\mu-\nu)\p_x+\p_u$ \\
%2*
\refstepcounter{tbn}\label{AeuBueufex2+xghhex2}\thetbn & $ue^u$ & $e^{p x^2+q x}$ & $e^{p x^2}$ & $e^{p x^2}$ &
$\p_t,\, (2p+q)t\p_t+\p_x-2p\p_u$ \\
%3
\refstepcounter{tbn}\label{AeuBenukf1h1}\thetbn & $e^{\nu u}+\varkappa$ & $1$ & $1$ & 1 &
$\p_t,\, \p_x,\, (\mu-2\nu)t\p_t+((\mu-\nu)x+\nu\varkappa t)\p_x+\p_u$ \\
%4
\refstepcounter{tbn}\label{AeuBuf1h1}\thetbn & $u$ & 1 & 1 & 1 &$\p_t,\, \p_x,\, t\p_t+(x-t)\p_x+\p_u$ \\
%5
\refstepcounter{tbn}\label{AeuB0ff1h}\thetbn a &
0 & $f^1(x)$ & 1 & 1  & $\p_t,\, t\p_t-\p_u,\, \alpha t\p_t+(\beta x^2+\gamma_1x+\gamma_0)\p_x+\beta x\p_u$ \\
\thetbn b &
1 & $|x|^p$ & 1  & $\varepsilon x|x|^{p}$ & $\p_t,\, x\p_x+(p+2)\p_u,\, e^{-\varepsilon(p+2)t}(\p_t-\varepsilon x\p_x)$\\
\thetbn b${}^*$ &
1 & $e^{x}$ & 1 & $\varepsilon e^{x}$ & $\p_t,\, \p_x+\p_u,\, e^{-\varepsilon t}(\p_t-\varepsilon\p_x)$ \\
\thetbn c & 1 & $x^{-2}$ & 1 & $\varepsilon x^{-1}$ & $\p_t,\, x\p_x,\, t\p_t-\varepsilon tx\p_x-\p_u$ \\
%6
\refstepcounter{tbn}\label{AeuB0f1h}\thetbn a & 0 & 1& 1 & 1 & $\p_t,\, t\p_t-\p_u,\, 2t\p_t+x\p_x,\, \p_x$ \\
\thetbn b & 1 & 1 & 1 & 1 & $\p_t, \,\p_x,\, t\p_t-t\p_x-\p_u,\, 2t\p_t+(x-t)\p_x$ \\
\thetbn c & 1 & $1$ & 1 & $\varepsilon x$ & $\p_t,\, x\p_x+2\p_u,\, e^{-\varepsilon t}\p_x,\,
e^{-2\varepsilon t}(\p_t-\varepsilon x\p_x)$ \\
\thetbn d & 0 & $x^{-3}$ & 1 & 1 & $ \p_t,\, t\p_t-\p_u,\, x\p_x-\p_u,\, x^2\p_x+x\p_u$ \\
\thetbn e & 1 & $x^{-3}$ & 1 & $x^{-2}$ & $\p_t,\, x\p_x-\p_u,\, e^t(\p_t-x\p_x),\, e^t(x^2\p_x+x\p_u)$ \\
%7
\hline
\end{tabular}
\end{center}
{\footnotesize
Here $(\mu,\,\nu)\in\{(0,\,1),\, (1,\,\nu)\}$, $\nu\ne\mu$ in cases~\ref{AeuBenufemxhex}, \ref{AeuBenufemxhex}${}^*$
%\ref{AeuBenufxmhxl}
and~\ref{AeuBenukf1h1};
$\mu=1$ and $\nu\ne1$ in the other cases;
$q\ne-1$ in case~\ref{AeuBenufemxhex}${}^*$ (otherwise it is subcase of the case~1.\ref{gcAaBafexh1}a$'$);
$\varepsilon=\pm1$ in cases~\ref{AeuBenufemxhex}, \ref{AeuB0ff1h}b--\ref{AeuB0ff1h}c and \ref{AeuB0f1h}e;
$p\not\in\{-3,-2,0\}$ in case~\ref{AeuB0ff1h}b;
$\alpha,\beta, \gamma_1, \gamma_0=\const$ and
\[
f^1(x)=\exp\left\{\int\frac{-3\beta x-2\gamma_1+\alpha}
 {\beta x^2+\gamma_1x+\gamma_0}\,dx\right \}.
\]
Case~\ref{AeuBenufemxhex}($q=-1$) is a subcase of case~1.\ref{AeuBenufemxhex}a${}'$.
\setcounter{casetran}{0}
Additional equivalence transformations:\\[1ex]
\refstepcounter{casetran}\thecasetran. \ref{AeuBenukf1h1}($\varkappa\ne0$) $\to$ \ref{AeuBenukf1h1}($\varkappa=0$):\quad
$\tilde t=t$, $\tilde x=x+\varkappa t$, $\tilde u=u$;
\\[1ex]
\refstepcounter{casetran}\thecasetran. \ref{AeuB0ff1h}b $\to$ \ref{AeuB0ff1h}a
($\beta=\gamma_0=0$, $\alpha=(p+2)\gamma_1$):\quad
$\tilde t=(e^{\varepsilon(p+2)t}-1)/(\varepsilon(p+2))$, $\tilde x=xe^{\varepsilon t}$, $\tilde u=u$;
\\[1ex]
\refstepcounter{casetran}\thecasetran. \ref{AeuB0ff1h}b${}^*$ $\to$ \ref{AeuB0ff1h}a
($\beta=\gamma_1=0$, $\alpha=\gamma_0$):\quad $\tilde t=e^{\varepsilon t}/\varepsilon$, $\tilde x=x+\varepsilon t$,
$\tilde u=u$;
\\[1ex]
\refstepcounter{casetran}\thecasetran. \ref{AeuB0ff1h}c $\to$ \ref{AeuB0ff1h}a
($\beta=\gamma_0=\alpha=0$):\quad $\tilde t=t$, $\tilde x=xe^{\varepsilon t}$, $\tilde u=u$;
\\[1ex]
\refstepcounter{casetran}\thecasetran.
\ref{AeuB0f1h}b$\to$\ref{AeuB0f1h}a:\quad $\tilde t= t$, $\tilde x=x+t$, $\tilde u=u$;
\\[1ex]
\refstepcounter{casetran}\thecasetran. \ref{AeuB0f1h}c$\to$\ref{AeuB0f1h}a:\quad
$\tilde t=e^{2\varepsilon t}/(2\varepsilon)$, $\tilde x=xe^{\varepsilon t}$, $\tilde u=u$;
\\[1ex]
\refstepcounter{casetran}\thecasetran.
\ref{AeuB0f1h}d$\to$\ref{AeuB0f1h}a:\quad $\tilde t= t\sign x$, $\tilde x=1/x,$ $\tilde u=u-\ln |x|$;
\\[1ex]
\refstepcounter{casetran}\thecasetran. \ref{AeuB0f1h}e$\to$\ref{AeuB0f1h}a:\quad $\tilde t=(e^{2t}t\sign x)/2$,
$\tilde x=e^{-t}/x,$ $\tilde u=u-t-\ln |x|$.
%%%%%%%%%%
\par}
\newpage
\setcounter{tbn}{0}

\begin{center}\footnotesize\renewcommand{\arraystretch}{1.2}\refstepcounter{tabul}
Table~\thetabul. Case of $A(u)=|u|^\mu$ \\[1ex]
\begin{tabular}{|l|l|c|c|c|c|l|}
\hline
\hfil N & $\hfil \mu$ & $B(u)$ & $f(x)$ & $g(x)$ & $h(x)$ & \hfil Basis of A$^{\rm max}$ \\
\hline
\refstepcounter{tbn}\label{AumB0fh}\thetbn & $\forall$ & $0$ & $\forall$ & 1 & 1 & $\p_t,\, \mu t\p_t-u\p_u$ \\
%1
\refstepcounter{tbn}\label{AumBunfxlhxg}\thetbn&
$\forall$ & $|u|^\nu$ & $|x|^p$ & 1 & $|x|^q$ & $\p_t,\, (\mu+p\mu-q\mu-p\nu-2\nu)t\p_t$ \\
&&&&&&$+(\mu-\nu)x\p_x+(q+1)u\p_u$ \\
%2
%\refstepcounter{tbn}\label{AumBunfelxhegx}
\thetbn${}^*$ &
$\forall$ & $|u|^\nu$ & $e^{px}$ & 1 & $\varepsilon e^{x}$ & $\p_t,\, (p\mu-p\nu-\mu)t\p_t+(\mu-\nu)\p_x+u\p_u$ \\
%2*
\refstepcounter{tbn}\label{aumbumlnufeax2+xghheax}\thetbn & $\forall$ & $|u|^\mu\ln|u|$ & $e^{p x^2+q x}$ &
$e^{p x^2}$ & $e^{p x^2}$ & $\p_t,\, (2\mu p+q)t\p_t+\p_x-2p u\p_u$ \\
%3
\refstepcounter{tbn}\label{Au-1B1fx-3ebxhx-2ebx}\thetbn & $\forall$ & 1 & $f^2(x)$ & 1 & $\varepsilon xf^2(x)$
& $\p_t,$ \\
&&&&&& $e^{\varepsilon t}(\p_t-\varepsilon((\mu+1)\beta x^2+x)\p_x-\varepsilon\beta xu\p_u)$ \\
%4
\refstepcounter{tbn}\label{a1bafeax2ghheax2}\thetbn & 0 & $\forall$ & $e^{p x^2}$ &
$e^{p x^2}$ & $e^{p x^2}$ & $\p_t,\, e^{-2p t}\p_x$ \\
%5
\refstepcounter{tbn}\label{a1bafexhghhgex}\thetbn & 0 & $\forall$ & $ e^{x+\gamma e^{x}}$
& $e^{\gamma e^{x}}$ & $e^{\gamma e^{x}}$ &
$\p_t,\, e^{-\gamma t}(\p_t-\gamma \p_x)$ \\
%6
\refstepcounter{tbn}\label{a1bufeax2+xghheax}\thetbn & 0 & $u$ & $e^{p x^2+x}$ &
$e^{p x^2}$ & $e^{p x^2}$ & $\p_t,\, t\p_t+\p_x-2p\p_u$ \\
%7
\refstepcounter{tbn}\label{AumBunkf1h1}\thetbn & $\forall$ & $|u|^\nu+\varkappa$ & $1$ &1 & $1$ & $\p_t,\, \p_x,$ \\
&&&&&& $(\mu-2\nu)t\p_t+((\mu-\nu)x+\nu\varkappa t)\p_x+u\p_u$ \\
%8
\refstepcounter{tbn}\label{AumBlnuf1h1}\thetbn & $\forall$ & $\ln|u|$ & $1$ &1 & $1$ &
$\p_t,\, \p_x,\, \mu t\p_t+(\mu x-t)\p_x+u\p_u$ \\
%9
\refstepcounter{tbn}\label{a1bufeax2ghheax2}\thetbn & 0 & $u$ & $e^{p x^2}$ & $e^{p x^2}$ & $e^{p x^2}$ &
$\p_t,\, e^{-2p t}\p_x,\, \p_x-2p\p_u$ \\
%10
\refstepcounter{tbn}\label{a1blnufeax2ghheax2}\thetbn & 0 & $\ln|u|$ & $e^{p x^2}$ & $e^{p x^2}$ & $e^{p x^2}$ &
$\p_t,\, e^{-2p t}\p_x,\, \p_x-2p u\p_u$ \\
%11
\refstepcounter{tbn}\label{AumB0ff3hh}\thetbn a & $\forall$ & 0 & $f^3(x)$ &1 & 1 &$\p_t,\, \mu t\p_t-u\p_u,$ \\
&&&&&& $\alpha t\p_t+((\mu+1)\beta x^2+\gamma_1x+\gamma_0)\p_x+\beta xu\p_u$ \\
%
%\refstepcounter{tbn}\label{AumB0fxph}\thetbn a & $\forall$ & $0$ & $|x|^p$ & 1 &
%$\p_t,\, (p+2)t\p_t+x\p_x,\, \mu t\p_t-u\p_u$ \\
%
\thetbn b & $\forall$ & $1$ & $|x|^p$ &1 & $\varepsilon x|x|^{p}$ &
$\p_t,\, \mu x\p_x+(p+2)u\p_u,\, e^{-\varepsilon(p+2)t}(\p_t-\varepsilon x\p_x)$ \\
%
%\refstepcounter{tbn}\label{AumB1fxlhxl1}
\thetbn b${}^*\!\!\!$ &
$\ne-1$ & 1 & $e^{x}$ &1 & $\varepsilon e^{x}$ & $\p_t,\, \mu\p_x+u\p_u,\, e^{-\varepsilon t}(\p_t-\varepsilon\p_x)$ \\
\thetbn c & $\ne-2$ & 1 & $x^{-2}$ &1 & $\varepsilon x^{-1}$ & $\p_t,\, x\p_x,\, \mu t\p_t-\varepsilon\mu tx\p_x-u\p_u$\\
%12
\refstepcounter{tbn}\label{A-65B1fx2hx2}\thetbn & $-6/5$ & $1$ & $x^2$ &1 & $x^2$ & $\p_t,\, 2t\p_t+2x\p_x-5u\p_u,$ \\
&&&&&& $t^2\p_t+(2tx+x^2)\p_x-5(t+x)u\p_u$ \\
%13
\refstepcounter{tbn}\label{AumB0f1h}\thetbn a & $\ne -4/3$ & 0 & 1 &1 & 1 &
$\p_t,\, \mu t\p_t-u\p_u,\, \p_x,\, 2t\p_t+x\p_x$\\
\thetbn b & $\ne -4/3$ & 1 &1 & 1 & 1 & $\p_t,\, \mu t\p_t-\mu t\p_x-u\p_u,\, \p_x,\, 2t\p_t+(x-t)\p_x$ \\
\thetbn c & $\ne -4/3$ & 1 & $1$ &1 & $\varepsilon x$ &
$\p_t,\, \mu x\p_x+2u\p_u,\, e^{-\varepsilon t}\p_x,\, e^{-2\varepsilon t}(\p_t-\varepsilon x\p_x)$ \\
\thetbn d & $\ne -4/3, -1$ & 0 & $|x|^{-\frac{3\mu+4}{\mu+1}}$ &1 & 1 &
$\p_t,\, \mu t\p_t-u\p_u,\, (\mu+2)t\p_t-(\mu+1)x\p_x,$ \\ &&&&&& $(\mu+1)x^2\p_x+xu\p_u $ \\
\thetbn e&$\ne -2,$  & 1 & $|x|^{-\frac{3\mu+4}{\mu+1}}$ &1 & $\varepsilon x|x|^{-\frac{3\mu+4}{\mu+1}}$ &
$\p_t,\, \mu(\mu+1)x\p_x-(\mu+2)u\p_u,$ \\
&$-4/3,-1$&&&&&$e^{\varepsilon\frac{\mu+2}{\mu+1}t}(\p_t-\varepsilon x\p_x),\,e^{\varepsilon t}((\mu+1)x^2\p_x+xu\p_u)$\\
\thetbn f & $-1$ & 0 & $e^x$ &1 & 1 & $ \p_t,\, t\p_t+u\p_u,\, \p_x-u\p_u,\,2t\p_t+x\p_x-xu\p_u$ \\
\thetbn g & $-1$ & $1$ & $e^{x}$ &1 & $\varepsilon e^{x}$ &
$\p_t,\, \p_x-u\p_u,\, (x+\varepsilon t-2)\p_x-(x+\varepsilon t)u\p_u,$ \\
&&&&&& $e^{-\varepsilon t}(\p_t-\varepsilon\p_x)$ \\
\thetbn h & $-2$ & $1$ & $x^{-2}$ &1 & $\varepsilon x^{-1}$ & $\p_t,\, x\p_x,\, 2t\p_t-2\varepsilon tx\p_x+u\p_u,$ \\
&&&&&& $e^{\varepsilon t}(x^2\p_x-xu\p_u)$\\
%14
\refstepcounter{tbn}\label{A-43B0f1h}\thetbn a & $-4/3$ & 0 & 1 &1 & 1 &
$\p_t,\, 4t\p_t+3u\p_u,\, \p_x,\, 2t\p_t+x\p_x,$ \\ &&&&&& $x^2\p_x-3xu\p_u$ \\
\thetbn b & $-4/3$ & 1 & 1 &1 & $1$ & $\p_t,\, 4t\p_t+4x\p_x-3u\p_u,\, 2t\p_t+(x-t)\p_x,$ \\
 &&&&&&$\p_x,\, (x+t)^2\p_x-3(x+t)u\p_u$ \\
\thetbn c & $-4/3$ & 1 & 1 &1 & $\varepsilon x$ & $\p_t,\, 2x\p_x-3u\p_u,\, e^{-\varepsilon t}\p_x,$ \\
&&&&&& $e^{-2\varepsilon t}(\p_t-\varepsilon x\p_x),\, e^{\varepsilon t}(x^2\p_x-3xu\p_u)$ \\
%15
\refstepcounter{tbn}\label{A1Buf1h1}\thetbn & 0 & $u$ & 1 &1 & $1$
& $ \p_t,\, \p_x,\, t\p_x-\p_u,\, 2t\p_t+x\p_x-u\p_u,$ \\
&&&&&& $t^2\p_t+tx\p_x-(tu+x)\p_u$ \\
%
%\thetbn b&$0$&$u$&$x^{-2}$&$\varepsilon x^{-1}$&
%$\p_t,\, x\p_x,\, \varepsilon tx\p_x-\p_u,\, 2t\p_t+x(\ln x+t)\p_x-u\p_u,$\\
%&&&&&$\varepsilon t^2\p_t+\varepsilon tx\ln x\p_x-(\varepsilon tu-t+\ln x)\p_u$ \\
%16
\hline
\end{tabular}
\end{center}
{\footnotesize
Here $\nu\ne\mu$; $\varepsilon=\pm1$;
$q\ne-1$ in case~\ref{AumBunfxlhxg}${}^*$ (otherwise it is subcase of the case~1.\ref{gcAaBafexh1}a$'$);
%$\beta\ne0$ in cases~\ref{AumBumlnuffhh},~\ref{A1Blnufh2hh};
%$\gamma_1\ne0$ in case~\ref{Au-1B1fx-3ebxhx-2ebx};
$p\ne-2,-(3\mu+4)/(\mu+1)$ in case~\ref{AumB0ff3hh}c;
$\alpha$, $\beta$, $\gamma_1$, $\gamma_0=\const$,
%$\varphi=\varphi(x)$ satisfies equation $\varphi_x=ce^{\alpha\varphi/\beta^2}|\varphi|^\gamma$
and
\[
f^2(x)=\exp\left\{\int\frac{-(3\mu+4)\beta x-3}
 {(\mu+1)\beta x^2+x}\,dx\right \},\quad
f^3(x)=\exp\left\{\int\frac{-(3\mu+4)\beta x-2\gamma_1+\alpha}
 {(\mu+1)\beta x^2+\gamma_1x+\gamma_0}\,dx\right \}.
\]
Additional equivalence transformations:\\[1ex]
\setcounter{casetran}{0}
\refstepcounter{casetran}\thecasetran.
\ref{AumBunkf1h1}($\varkappa\ne0$) $\to$ \ref{AumBunkf1h1}($\varkappa=0$):\quad
$\tilde t=t$, $\tilde x=x+\varkappa t$, $\tilde u=u$;
\\[1ex]
\refstepcounter{casetran}\thecasetran.
\ref{AumB0ff3hh}b $\to$ \ref{AumB0ff3hh}a ($\beta=\gamma_0=\alpha=0$), \ref{AumB0f1h}e $\to$ \ref{AumB0f1h}a:\quad
$\tilde t=(e^{(p+2)t}-1)/(p+2)$, $\tilde x=xe^t$, $\tilde u=u$;
\\[1ex]
\refstepcounter{casetran}\thecasetran. \ref{AumB0ff3hh}b${}^*$ $\to$ \ref{AumB0ff3hh}a
($\beta=\gamma_1=0$, $\alpha=\gamma_0$):\quad
$\tilde t=(e^{\varepsilon(p+2)t}-1)/(\varepsilon(p+2))$, $\tilde x=xe^{\varepsilon t}$, $\tilde u=u$;
\\[1ex]
\refstepcounter{casetran}\thecasetran. \ref{AumB0ff3hh}c $\to$ \ref{AumB0ff3hh}a ($\beta=\gamma_0=\alpha=0$),
\ref{AumB0f1h}h $\to$ \ref{AumB0f1h}a:\quad
$\tilde t=t$, $\tilde x=xe^{\varepsilon t}$, $\tilde u=u$;
\\[1ex]
\refstepcounter{casetran}\thecasetran.
\ref{AumB0f1h}b $\to$ \ref{AumB0f1h}a, \ref{A-43B0f1h}b $\to$ \ref{A-43B0f1h}a:\quad
$\tilde t=t$, $\tilde x=x-t$, $\tilde u=u$;
\\[1ex]
\refstepcounter{casetran}\thecasetran.
\ref{AumB0f1h}c $\to$ \ref{AumB0f1h}a, \ref{A-43B0f1h}c $\to$ \ref{A-43B0f1h}a:\quad
$\tilde t=e^{2\varepsilon t}/(2\varepsilon)$, $\tilde x=xe^{\varepsilon t}$, $\tilde u=u$;
\\[1ex]
\refstepcounter{casetran}\thecasetran.
\ref{AumB0f1h}d $\to$ \ref{AumB0f1h}a: $\tilde t=t$, $\tilde x=-1/x$, $\tilde u=|x|^{-\frac1{1+\mu}}u$;
\\[1ex]
\refstepcounter{casetran}\thecasetran.
\ref{AumB0f1h}f $\to$ \ref{AumB0f1h}a($\mu=-1$): $\tilde t=t$, $\tilde x=x$, $\tilde u=e^{x}u$.
\\[1ex]
\refstepcounter{casetran}\thecasetran.
\ref{AumB0f1h}g $\to$ \ref{AumB0f1h}a($\mu=-1$):
  $\tilde t=e^{\varepsilon t}/\varepsilon$, $\tilde x=x+\varepsilon t$, $\tilde u=e^{x+\varepsilon t}u$;\\[1ex]
\par
}

\end{appendix}

\end{document}